\documentclass[a4paper, 11pt]{article}
\usepackage{
  jheppub
} 
\usepackage{lineno}

\usepackage{booktabs} 

\arxivnumber{2602.09354} 

\title{Hadronic decay branching ratio measurements of the Higgs boson at future colliders using the Holistic Approach}

\author[a,b]{Jianfeng Jiang}
\author[a]{Yongfeng Zhu}
\author[a,b]{Chao Yang}
\author[a,1]{Manqi Ruan\note{Corresponding author.}}

\affiliation[a]{Institute of High Energy Physics, Chinese Academy of Sciences, Beijing 100049, China}
\affiliation[b]{University of Chinese Academy of Sciences (UCAS), Beijing 100049, China}

\emailAdd{manqi.ruan@ihep.ac.cn}

\abstract{
Accurately measuring the properties of the Higgs boson is one of the primary
physics objectives of the high-energy frontier.
By incorporating the inclusive information of all reconstructed particles to identify the
signal events, referred to as the holistic approach,
we estimate the relative statistical uncertainty for
the Higgs hadronic decay modes $H \to b\bar{b}$, $c\bar{c}$, $gg$, $WW^{*}\to 4q$, and $ZZ^{*}\to 4q$ at the Circular Electron–Positron Collider (CEPC) operating as a Higgs factory with
an integrated luminosity of 21.6~ab$^{-1}$. 
In the $Z(\mu^{+}\mu^{-})H$ and $Z(\nu\overline{\nu})H$ channels, the relative statistical uncertainties for these decay modes are projected to range from 0.36\% to 5.21\% and 0.16\% to 2.52\%, respectively. 
Compared to the CEPC Snowmass results, the holistic approach boosts the measurement precision by a factor of two to four. 
The scaling behavior, specifically the dependence of the anticipated accuracy on the training
dataset size, is observed and analyzed. 
The precision of these leading Higgs decay modes, especially the $H\to b\bar{b}$ mode, is asymptotically approaching the statistical limit. 
The scaling behavior could also be applied to monitor the robustness and to quantify the uncertainties of the holistic approach.
}

\begin{document}
\maketitle
  \flushbottom

  \section{Introduction}
  \label{sec:introduction}

  After the discovery of the Higgs boson at the Large Hadron Collider (LHC) in 2012~\cite{ATLAS:2012yve,CMS:2012qbp},
  a primary objective of high-energy physics has been the search for new physics via the Higgs portal, especially through the precise measurement of the Higgs boson couplings.
  The hadronic decays of the Higgs boson account for a cumulative branching ratio of approximately 80\%, making them critical for precise Higgs property measurements.
  Those measurements at the LHC are significantly limited due to the overwhelming QCD backgrounds and severe pile-up conditions.
  Meanwhile, identified as the highest-priority future collider facility~\cite{European:2720131,deBlas:2025gyz}, the electron-positron Higgs factory offers a clean collision environment ideal for exploring these hadronic decay modes.
  Prominent proposals include the Circular Electron–Positron
  Collider (CEPC)\cite{CEPCStudyGroup:2018rmc,CEPCStudyGroup:2018ghi,CEPCStudyGroup:2023quu,CEPCStudyGroup:2025kmw},
  the Future Circular Collider (FCC-ee)\cite{Agapov:2022bhm}, the International
  Linear Collider (ILC)~\cite{ILC:2013jhg}, the Compact Linear Collider (CLIC)~\cite{linssen2012physicsdetectorsclicclic} and many others such as LEP3~\cite{Anastopoulos:2025jyh}.

  Artificial intelligence (AI) has significantly advanced
  data analysis techniques in high-energy physics~\cite{DelVecchio:2025gzw}. Jet origin identification (JOI)~\cite{Liang:2023wpt}
  classifies jets according to their underlying parton flavor, while one-to-one
  reconstruction techniques~\cite{Wang:2024eji} associate detector hits with individual
  reconstructed particles to identify their species. More recently, an Advanced Color-Singlet
  Identification (ACSI) algorithm~\cite{Zhu:2025eoe} have been developed to directly associate final-state particles with their parent bosons.

  To exploit the physics potential of the electron-positron Higgs factory, we adopt the holistic approach to the measurements of leading Higgs hadronic decay modes. 
  The holistic approach is an end-to-end process that treats each collision event as an ensemble of reconstructed particles, it reads in the inclusive particle-level information and infers the event type according to its training.
  By utilizing high-dimensional input data, this approach incorporates approximately two orders of magnitude more input information compared to traditional method.
  We also observe the scaling behavior, which describes the evolution of performance with changing training data size.
  The scaling behavior is then applied to the analysis optimization, AI model behavior monitoring, and uncertainty control.
  
  
  This paper is organized as follows: Section~\ref{sec:methodology} introduces
  the detector model, software framework, AI algorithm, and simulated data samples
  used in this analysis. Section~\ref{sec:higgs_hadronic_decay_measurements}
  presents the analyses in the $Z(\mu^{+}\mu^{-})H$ and $Z(\nu\bar{\nu})H$
  channels and the extrapolated results. Section~\ref{sec:scaling_with_data_size} discusses
  the observed scaling behavior and the generator robustness of the holistic approach. Section~\ref{sec:discussion_and_summary}
  provides a summary and outlook.

  \section{Detector, Samples, and Methods}
  \label{sec:methodology}

\begin{figure*}[htbp]
    \centering
    \hfill
    \includegraphics[width=0.48\textwidth]{
      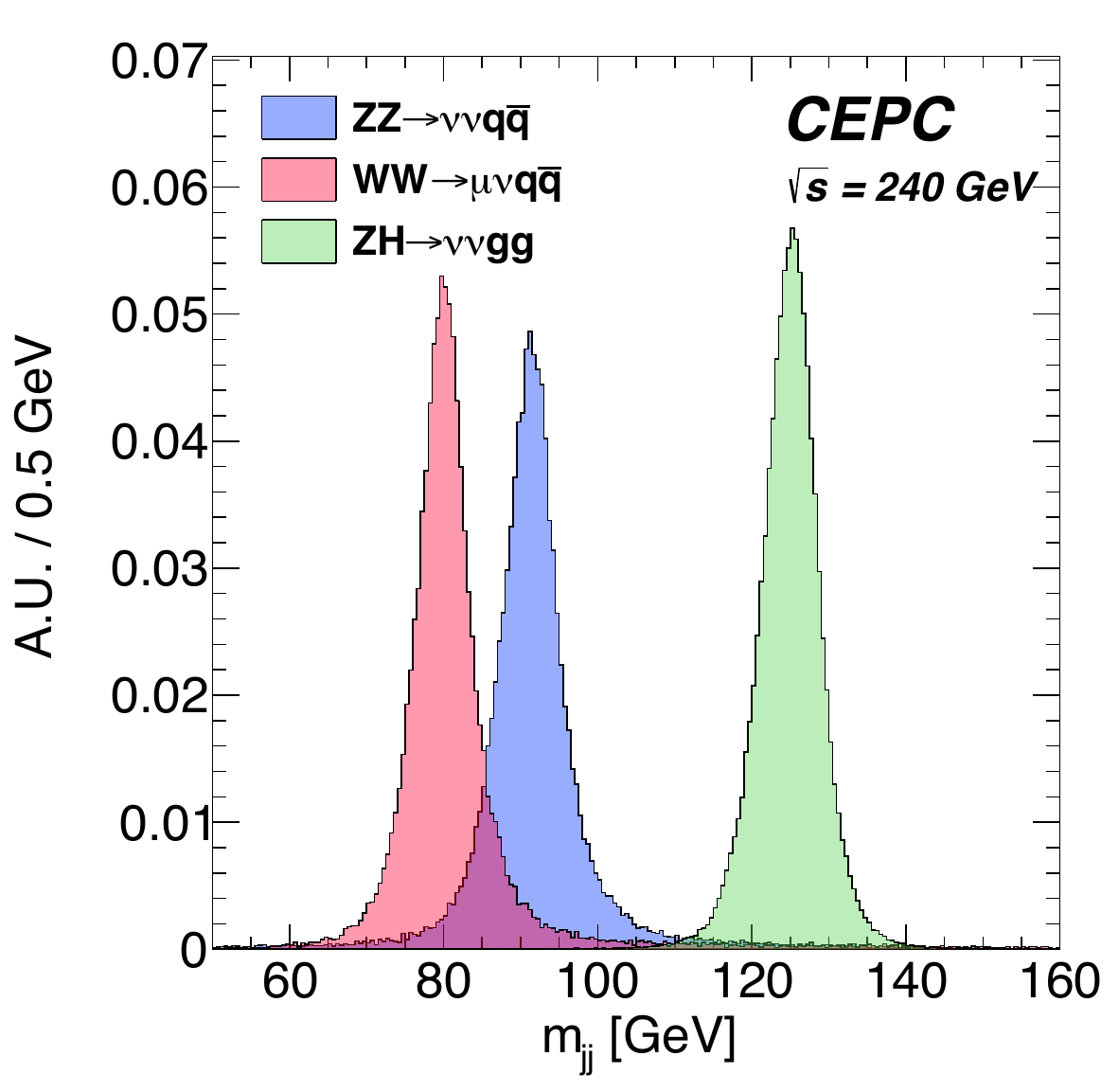
    }%
    \hfill
    \includegraphics[width=0.48\textwidth]{
      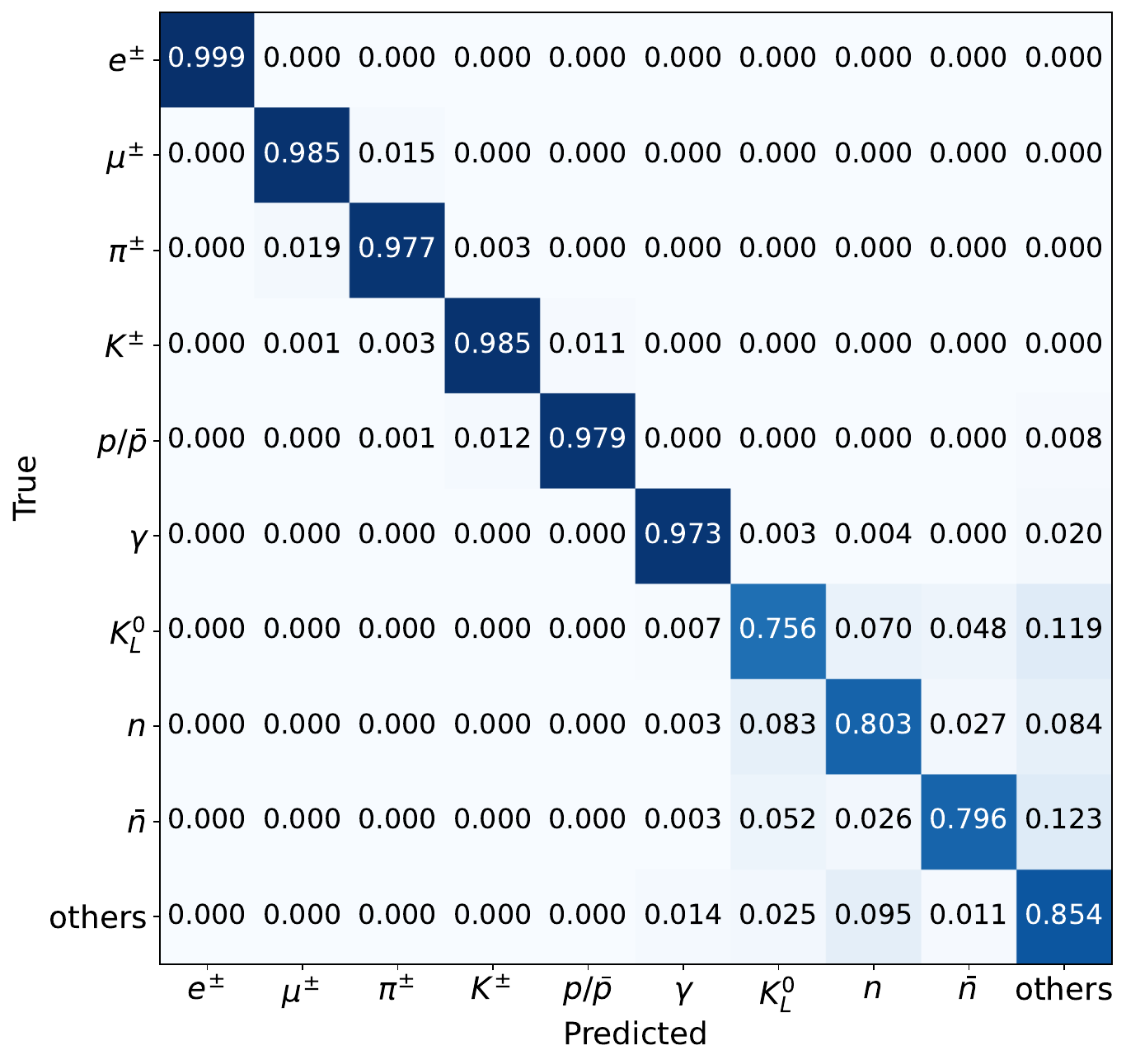
    }%
    \caption{Overview of the AURORA detector performance.
Left: The reconstructed invariant mass distributions for hadronic systems in $WW \to \mu\nu q\bar{q}$, $ZZ \to \nu\nu q\bar{q}$, and $ZH \to \nu\nu gg$ processes at $\sqrt{s}=240$ GeV. 
Right: Migration matrix for particle identification (PID) reconstruction, illustrating the identification efficiency for charged particles, photons, and neutral hadrons.
The detector concept achieves near-universal particle identification performance.}
    \label{fig:cepc_detector}
  \end{figure*}

  This study is based on the CEPC operating as a Higgs factory at a center-of-mass energy of 240 GeV, using the AURORA detector model~\cite{Wang:2024eji}.
  AURORA is a particle-flow–oriented detector design evolved from the CEPC Conceptual Design Report (CDR) baseline detector~\cite{CEPCStudyGroup:2018ghi}.
  It features high-granularity electromagnetic and hadronic calorimeters (ECAL and
  HCAL), a high-precision tracking system with a low material budget, a high-resolution
  vertex detector, and a large solenoidal magnet enclosing the calorimeters. In
  addition, both ECAL and HCAL are assumed to provide a per-cell time resolution of $O(100)$ ps. 
  The critical performance of AURORA is summarized in Figure~\ref{fig:cepc_detector}. 
  AURORA achieves a BMR of 2.7\%, which is a critical figure of merit representing the relative uncertainties on the reconstructed invariant mass of heavy bosons especially the Higgs boson. 
  Equipped with high-precision and high-granularity 5D calorimetry, AURORA offers efficient identification of the reconstructed particles, 97-100\% simultaneous efficiencies for charged particles and photon, and 75-80\% for neutral hadrons.

  \begin{table}
    \centering
    \begin{tabular}{lrr}
      \toprule Channel                        & Cross section ($\times$ BR) [fb] & Expected yield ($21.6\,\text{ab}^{-1}$) \\
      \midrule $Z(\mu^{+}\mu^{-})H(b\bar{b})$ & 3.8                              & $8.3 \times 10^{4}$                   \\
      $Z(\mu^{+}\mu^{-})H(c\bar{c})$          & 0.19                             & $4.2 \times 10^{3}$                   \\
      $Z(\mu^{+}\mu^{-})H(gg)$                & 0.57                             & $1.2 \times 10^{4}$                   \\
      $Z(\mu^{+}\mu^{-})H(WW^{*})$            & 0.65                             & $1.4 \times 10^{4}$                   \\
      $Z(\mu^{+}\mu^{-})H(ZZ^{*})$            & 0.086                            & $1.8 \times 10^{3}$                   \\
      $Z(\mu^{+}\mu^{-})Z(q\bar{q})$          & 51                               & $1.1 \times 10^{6}$                   \\
      \midrule $Z(\nu\bar{\nu})H(b\bar{b})$   & 23                               & $4.9 \times 10^{5}$                   \\
      $Z(\nu\bar{\nu})H(c\bar{c})$            & 1.2                              & $2.5 \times 10^{4}$                   \\
      $Z(\nu\bar{\nu})H(gg)$                  & 3.4                              & $7.3 \times 10^{4}$                   \\
      $Z(\nu\bar{\nu})H(WW^{*})$              & 3.8                              & $8.3 \times 10^{4}$                   \\
      $Z(\nu\bar{\nu})H(ZZ^{*})$              & 0.51                             & $1.1 \times 10^{4}$                   \\
      $W(l\nu)W(q\bar{q})$                    & 1200                             & $2.6 \times 10^{7}$                   \\
      $Z(\nu\bar{\nu})Z(q\bar{q})$            & 150                              & $3.2 \times 10^{6}$                   \\
      $q\bar{q}$                              & 54000                            & $1.2 \times 10^{10}$                  \\
      \bottomrule
    \end{tabular}
    \caption{Expected yields and cross section($\times \text{BR}$) for $Z(\mu^{+}
    \mu^{-})H$, $Z(\nu\bar{\nu})H$ and main background processes at 240~GeV
    center-of-mass energy with an integrated luminosity of $21.6\,\text{ab}^{-1}$.}
    \label{tab:expected_yields}
  \end{table}

  The signal samples, including $Z(\mu^{+}\mu^{-})H$ and $Z(\nu\bar{\nu})H$ with the Higgs boson decaying into five dominant hadronic modes ($H \to b\bar{b}$, $c\bar{c}$, $gg$, $WW^{*}$, $ZZ^{*}$), as well as the dominant Standard Model background processes ($e^{+}e^{-}\to ZZ$, $WW$, and $q\bar{q}$),
  were generated at Leading Order (LO) using MadGraph5\_aMC@NLO~\cite{Alwall:2014hca}.
  The \texttt{isronlyll} parameterization~\cite{Frixione:2021zdp} was employed to simulate Initial State Radiation (ISR), incorporating Leading-Logarithmic (LL) QED electron structure functions (PDFs) to model collinear photon radiation.
  Parton showering and hadronization were simulated with Pythia8~\cite{Bierlich:2022pfr} using the default Monash 2013 tune.
  To ensure consistency and avoid double-counting, ISR generation in Pythia8 was disabled, given that these effects were already treated by the matrix element generator.
  Additionally, multi-parton interactions (MPI) were turned off, reflecting the clean nature of electron-positron collisions.
  Using the fast simulation tool Delphes~\cite{deFavereau:2013fsa}, we mimic the performance of the AURORA detector~\cite{Wang:2024eji}, which features a BMR of approximately 3\% and ideal particle identification (PID) for charged particles.
  The cross sections and expected
  event yields for all signal and background processes, normalized to an integrated
  luminosity of $21.6~\mathrm{ab}^{-1}$, are summarized in Table~\ref{tab:expected_yields}.

  The holistic approach can be realized using different deep learning architectures.
  In this work, it is implemented with ParticleNet(PN)~\cite{Qu:2019gqs}, whose core building block is the EdgeConv operation that aggregates features from each particle and its $k$ nearest neighbors.
  By stacking multiple EdgeConv blocks, the network performs message passing to learn
  local and global correlations, enabling the extraction of high-level
  representations directly from the particle cloud. The EdgeConv layers are followed
  by a channel-wise global average pooling layer and two fully connected layers,
  with a softmax activation used to produce event-level classification scores. The
  input features provided to the network encode comprehensive information from reconstructed
  particles, including:
  \begin{itemize}
    \item \textit{Geometric coordinates}: The angular variables $\Delta\eta$ and
      $\Delta\phi$, defined relative to a reference axis.
      \footnote{ To explicitly
      capture the global topology of the $ZH$ process, a reference axis is
      introduced for coordinate transformation. For the $\mu^{+}\mu^{-}H$ channel,
      the reference axis is defined by the recoil system (the Higgs candidate), $p
      _{\mathrm{recoil}}= p_{\mathrm{initial}}- p_{\mu\mu}$. 
      For the
      $\nu\bar{\nu}H$ channel, it is defined by the missing momentum vector (the
      $Z$ boson candidate),
      $p_{\mathrm{missing}}= p_{\mathrm{initial}}- p_{\mathrm{visible}}$. These coordinates
      are used to construct the dynamic $k$-nearest-neighbor graph.
      }

    \item \textit{Kinematic variables}: The transverse momentum ($p_{T}$) and energy
      ($E$), together with their normalized fractions ($p_{T}/p_{T,\mathrm{evt}}$
      and $E/E_{\mathrm{evt}}$). The angular distance $\Delta R = \sqrt{\Delta\eta^{2}+
      \Delta\phi^{2}}$ and the particle charge are also included.

    \item \textit{Impact parameters}: The transverse ($d_{0}$) and longitudinal ($d
      _{z}$) impact parameters of the track, which are essential for resolving displaced
      vertices from heavy-flavor ($b$ and $c$) decays.

    \item \textit{Particle identification}: One-hot encoded labels for the seven reconstructed particle species:
      $\mu^{\pm}, e^{\pm}, \gamma, \pi^{\pm}, K^{\pm}, p^{\pm}$, and neutral hadrons.
  \end{itemize}

  The output of the network consists of a set of event-level classification scores, normalized via a softmax function, representing the probability of the event belonging to each signal or background category.

  For the $\mu^+\mu^- H$ channel, the model is trained on a dataset of $10^6$ events.
  A separate validation dataset of $2\times10^{5}$ events is employed for hyperparameter tuning and model checkpointing to mitigate overfitting.
  The final performance metrics are evaluated on an independent inference dataset of $8\times 10^{5}$ events, ensuring the statistical robustness of the results.
  For the $\nu\bar{\nu}H$ channel, the model is trained on a dataset of $10^5 \sim 10^6$ events, with
  $2\times10^{4}$ events allocated for validation and an independent inference dataset of $2\times10^{4}$ events.
  For both channels, the training process is carried out over 30 epochs.

  \section{Higgs Hadronic Decay Measurements}
  \label{sec:higgs_hadronic_decay_measurements}

  For the measurement of different Higgs decay processes in $\mu^{+}\mu^{-}H$ and $\nu\bar{\nu}H$ channels, we use the relative statistical uncertainty to evaluate the measurement performance.
  The relative statistical uncertainty is calculated as $\Delta \mu / \mu = \sqrt{S + B } / S$,
  where $S$ and $B$ denote the expected signal and background yields surviving the selection, respectively.

  \subsection{$\mu^+\mu^-H$ Channel}
  \label{sec:mmh}
  
  In the $\mu^{+}\mu^{-}H$ channel, the presence of two high-momentum
  muons allows for strong suppression of Standard
  Model backgrounds. However, the semileptonic $ZZ$ production remains the dominant background source~\cite{Ma:2024qoa}.
  We apply the holistic approach to directly discriminate among the different Higgs
  decay processes and the semileptonic $ZZ$ background. A training dataset of
  $10^{6}$ events is used for each process in a multi-class classification setup.

  \begin{figure*}[htbp]
    \centering
    \includegraphics[width=0.49\linewidth]{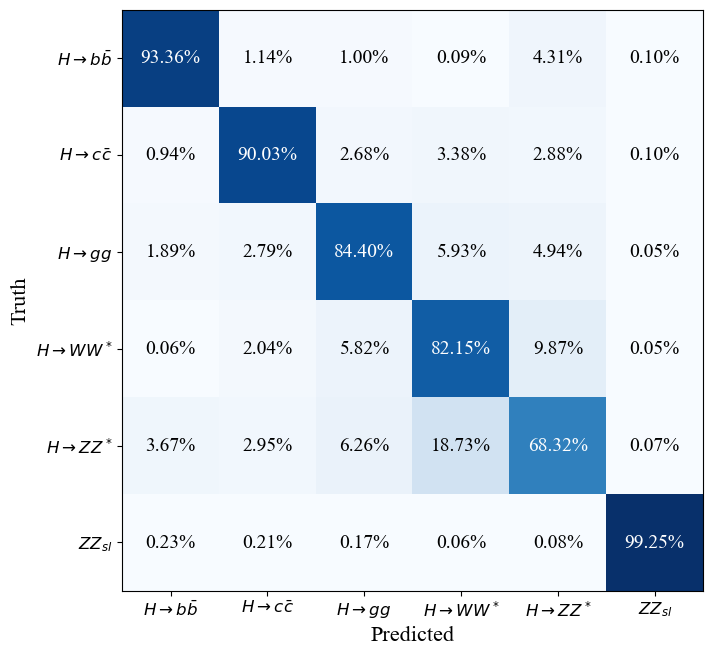}
    \hfill
    \includegraphics[width=0.49\linewidth]{
      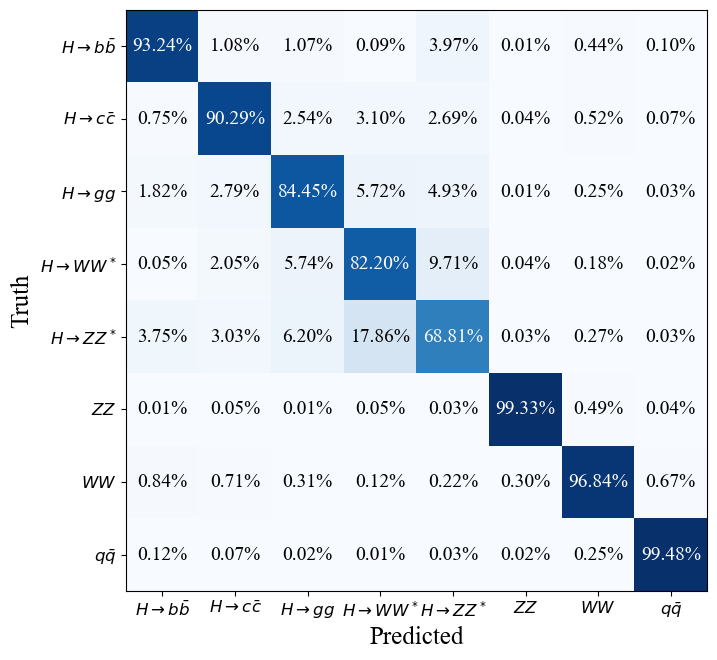
    }
    \caption{Migration matrix for the $\mu^{+}\mu^{-}H$ (left) and $\nu\bar{\nu}H$ (right) classification task
    using the holistic approach, trained on 1M events. Matrix elements are
    normalized to unity in each row.}
    \label{fig:Confusion_Matrix}
  \end{figure*}

  \begin{figure*}[htbp]
    \centering
    \includegraphics[width=0.48\textwidth]{
      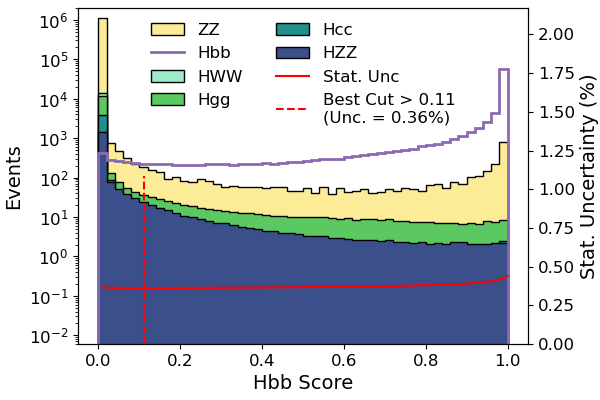
    }
    \includegraphics[width=0.48\textwidth]{
      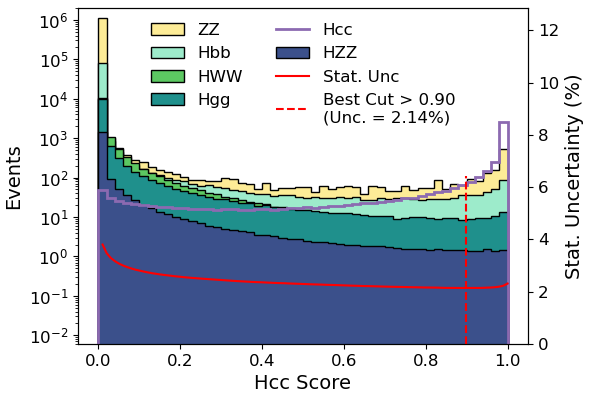
    }
    \includegraphics[width=0.48\textwidth]{
      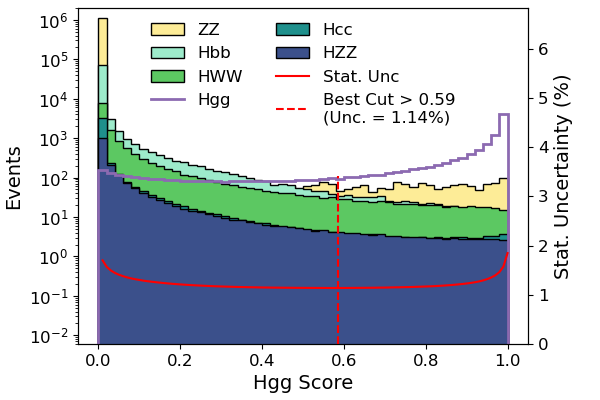
    }
    \includegraphics[width=0.48\textwidth]{
      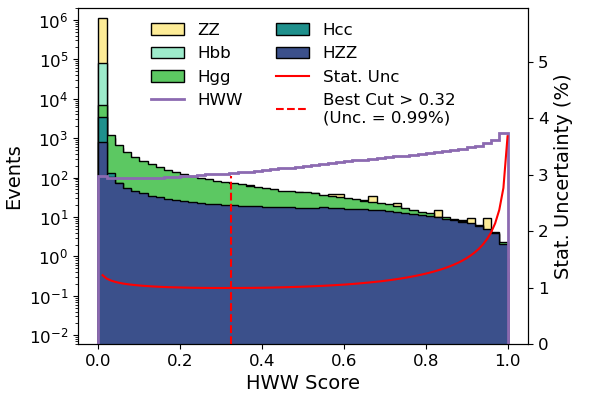
    }
    \includegraphics[width=0.48\textwidth]{
      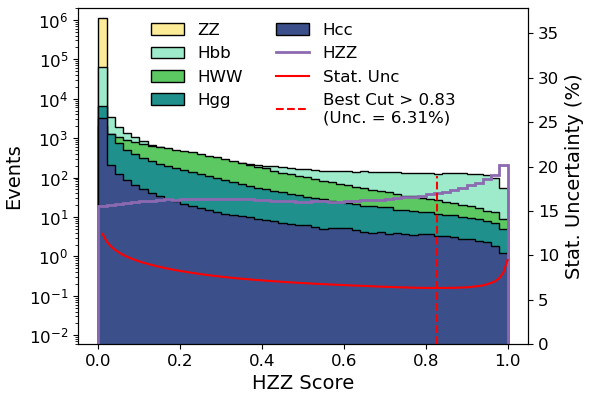
    }
    \caption{ Score distributions and relative statistical uncertainties for $H \to b\bar{b}$,
    $H \to c\bar{c}$, $H\to gg$, $H\to WW^{*}$, and $H \to ZZ^{*}$ processes in the
    $\mu^{+}\mu^{-}H$ analysis using the holistic approach trained on 1M events.
    The purple curve denotes the signal, while the colored histograms show the backgrounds.
    The red solid curve gives the statistical uncertainty versus the score cut, and
    the vertical red dashed line marks the optimal score cut. }
    \label{fig:mumuH_score_and_uncertainty}
  \end{figure*}

  The holistic approach takes the inclusive reconstructed particles as input, with a total dimensionality easily exceeding $10^3$, and simultaneously infers the probability scores for all different event types.
  By assigning each event to the category with the highest inferred score, we construct the migration matrix shown in Figure~\ref{fig:Confusion_Matrix}.
  Each row represents the truth-level process, while each column corresponds to the model's prediction, with each row normalized to unity.
  The diagonal elements represent the correct classification efficiencies. 
  A pronounced diagonal and symmetric\footnote{The migration matrices in Figure~\ref{fig:Confusion_Matrix} are symmetric except for the misidentification between $H\to WW^*$ and $H\to ZZ^*$, where $P(ZZ \to WW)$ is roughly two times larger than $P(WW\to ZZ)$. A plausible explanation of this phenomenon is the flavor structure of $H\to WW^* / ZZ^* \to 4q$ decay, as roughly half of the $H\to ZZ^* \to4q$ has $b$-quarks while $H\to WW^*$ has almost no $b$-quark in the $4q$ final states. This phenomenon is also observed in studies~\cite{DelVecchio:2025gzw,Ma:2024qoa}.}
  structure is observed. 
  The model exhibits strong performance on topologically distinct signatures.
  The dominant Standard Model background, $Z(\mu^{+}\mu^{-})Z(q\bar{q})$, achieves a high classification efficiency of over 99\%,
  indicating that it can be effectively suppressed in event selection.
  The Higgs decay processes achieve the highest efficiency of 93.4\% for the $H\to b\bar{b}$ channel, followed by 90.0\% $H \to c\bar{c}$ and 84.4\% for $H \to gg$.
  The off-diagonal elements quantify the misidentification rates among different channels. Similar behavior is also observed in the migration of the $\mu^+\mu^- H$ channel.

  To extract the signal and evaluate the measurement precision for each specific decay mode, we scan the threshold on its corresponding classification score. The optimal working point is determined by maximizing the product of signal efficiency and purity.
  The relative statistical uncertainty, as defined at the beginning of this section, is calculated at this optimized cut value.
  Figure~\ref{fig:mumuH_score_and_uncertainty} presents the best cut and the score distribution for the Higgs decay modes.
  Depending on the signal topology and background contamination, the optimal thresholds vary significantly.
  The $H \to b\bar{b}$ process requires only a loose threshold (score $> 0.11$), whereas the $H \to c\bar{c}$ process demands a strict cut
  (score $> 0.90$) to effectively suppress the background.
  The optimized statistical uncertainties are 0.36\% for $H\to b\bar{b}$, 2.14\% for $H\to c \bar{c}$, 1.14\% for $H\to gg$ and 0.99\% for $H\to WW^{*}$.
  
  For the $H\to ZZ^*$ channel, an anticipated relative statistical uncertainty of 6.31\% is achieved. 
  To fully exploit the classification information, we implemented a categorization strategy.
  $H \to b\bar{b}$ and $H\to WW^*$ are the primary backgrounds for the $H\to ZZ^*$ channel, we categorized the candidate events into four independent regions based on their inferred background scores and applied the optimal cut within each region. The detailed methodology is presented in Appendix~\ref{appendix:categorization_strategy}.

  This categorization strategy yields an approximate 20\% improvement for the $H\to ZZ^*$ measurement.
  We also applied this binned score method to the other hadronic decay processes.
  Table~\ref{tab:mmh_result} summarizes the projected statistical uncertainties before and after this optimization, alongside their respective statistical limits.
  \begin{table}[htbp]
    \centering
    \begin{tabular}{lccccc}
      \toprule  & $H \to b \bar{b}$ & $H \to c \bar{c}$ & $H \to g g$   & $H \to W W^*$ & $H \to Z Z^*$     \\
      \midrule 
       Score cut  & $0.36 \%$    & $2.14 \%$    & $1.14 \%$    & $0.99 \%$    & $6.31 \%$       \\
       Binned score cut       & $0.36 \%$    & $1.99 \%$    & $1.08 \%$    & $0.96 \%$    & $5.21 \%$       \\
       \midrule
      Stat. limit           & $0.35 \%$    & $1.55 \%$    & $0.90 \%$    & $0.85 \%$    & $2.33 \%$       \\
      \bottomrule
    \end{tabular}
    \caption{Projected relative statistical precision for the Higgs hadronic decay processes in the $Z(\mu^+\mu^-)H$ channel with an integrated luminosity of $21.6~\text{ab}^{-1}$.
    The table compares the results from the optimal score cut and the binned optimal score cut against the statistical limits.}
    \label{tab:mmh_result}
  \end{table}

  \subsection{$\nu\bar{\nu}H$ Channel}

  For the $\nu\bar{\nu}H$ channel, the classification performance exhibits a pattern highly similar to that of the $\mu^+\mu^-H$ analysis, as shown in the migration matrix (Figure~\ref{fig:Confusion_Matrix}).
  While the migration matrix shows decent separation power, the background cross sections in the $\nu\bar{\nu}H$ channel remain significantly large.
  The expected yield of the $q\bar{q}$ background alone exceeds $10^{10}$ events, dwarfing the signal by several orders of magnitude, see Table~\ref{tab:expected_yields}.
  To mitigate this immense background contamination, a pre-selection based on kinematic variables is applied, see Table~\ref{tab:scaled_cut_flow_scientific}. 
  At the cost of 5-20\% of signal efficiency, this pre-selection suppresses the SM backgrounds by nearly three orders of magnitudes.
  \begin{table}[htbp]
    \centering
    \small
    \begin{tabular}{lccccc}
      \toprule Channel                      
      & \begin{tabular}{c} Expected\\Events \end{tabular}     
      & \begin{tabular}{c}$p_{T,evt}> 10$\\GeV  \end{tabular} 
      & \begin{tabular}{c}$60 < M_{miss}$\\$< 140$~GeV \end{tabular} 
      & \begin{tabular}{c}$105 < M_{inv}$\\$< 160$~GeV \end{tabular} 
      & Eff (\%) \\
      \midrule $Z(\nu\bar{\nu})H(b\bar{b})$ & $4.9 \times 10^{5}$   & $4.77 \times 10^{5}$ & $4.63 \times 10^{5}$ & $3.93 \times 10^{5}$ & 80.20    \\
      $Z(\nu\bar{\nu})H(c\bar{c})$          & $2.5 \times 10^{4}$   & $2.41 \times 10^{4}$ & $2.38 \times 10^{4}$ & $2.20 \times 10^{4}$ & 89.29    \\
      $Z(\nu\bar{\nu})H(gg)$                & $7.3 \times 10^{4}$   & $7.13 \times 10^{4}$ & $7.08 \times 10^{4}$ & $6.94 \times 10^{4}$ & 95.30    \\
      $Z(\nu\bar{\nu})H(WW^{*})$            & $8.3 \times 10^{4}$   & $8.12 \times 10^{4}$ & $8.08 \times 10^{4}$ & $7.91 \times 10^{4}$ & 95.08    \\
      $Z(\nu\bar{\nu})H(ZZ^{*})$            & $1.1 \times 10^{4}$   & $1.07 \times 10^{4}$ & $1.06 \times 10^{4}$ & $1.01 \times 10^{4}$ & 92.09    \\
      \midrule $W(l\nu)W(q\bar{q})$         & $2.6 \times 10^{7}$   & $2.52 \times 10^{7}$ & $5.84 \times 10^{6}$ & $3.73 \times 10^{6}$ & 14.12    \\
      $Z(\nu\bar{\nu})Z(q\bar{q})$          & $3.2 \times 10^{6}$   & $3.18 \times 10^{6}$ & $3.00 \times 10^{6}$ & $5.99 \times 10^{4}$ & 1.83     \\
      $q\bar{q}$                            & $1.2 \times 10^{10}$  & $6.47 \times 10^{8}$ & $4.56 \times 10^{8}$ & $1.63 \times 10^{7}$ & 0.14     \\
      \bottomrule
    \end{tabular}
    \caption{
    The cut flow of pre-selection, normalized to an integrated luminosity of
    $21.6\,\text{ab}^{-1}$ at 240~GeV.
    $M_{miss}$ and $M_{inv}$ represent the total missing mass and the invariant mass of all visible particles, respectively, while the last column shows the overall pre-selection efficiency.
    }
    \label{tab:scaled_cut_flow_scientific}
  \end{table}

  \begin{figure*}[htbp]
    \centering
    \includegraphics[width=0.48\textwidth]{
      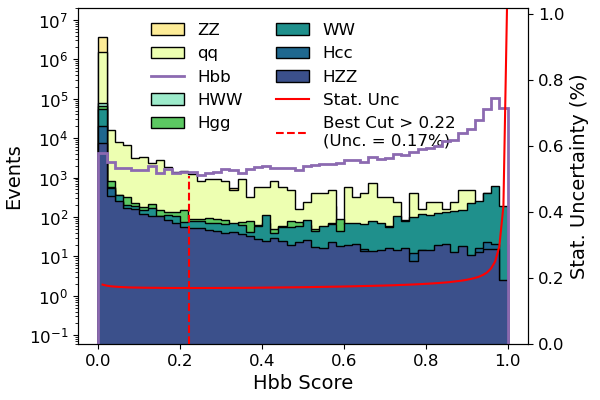
    }
    \includegraphics[width=0.48\textwidth]{
      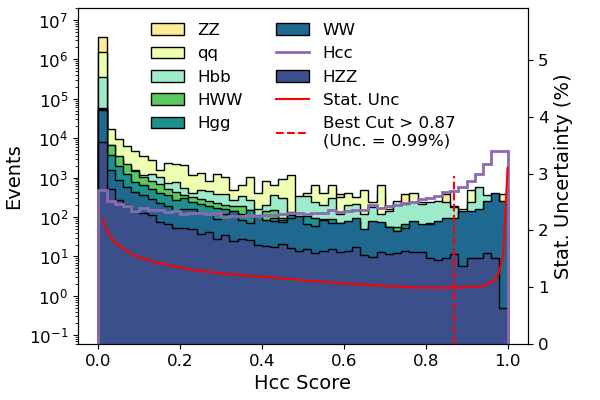
    }
    \includegraphics[width=0.48\textwidth]{
      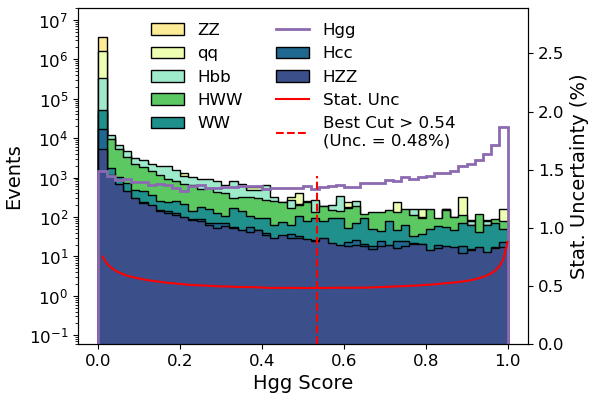
    }
    \includegraphics[width=0.48\textwidth]{
      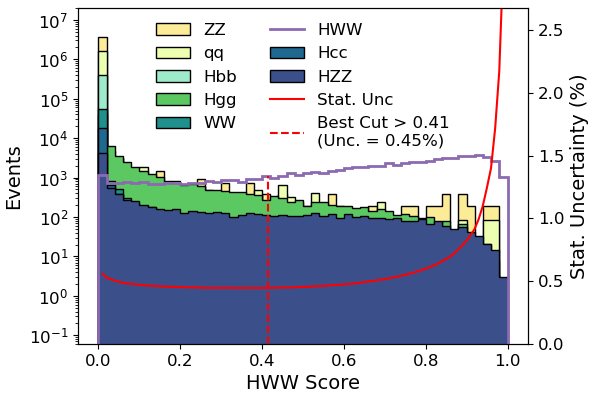
    }
    \includegraphics[width=0.48\textwidth]{
      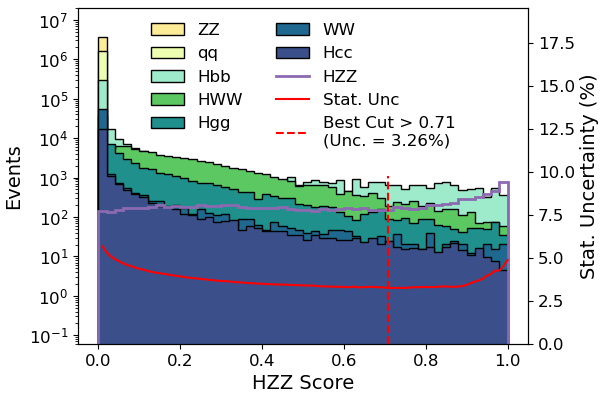
    }
    \caption{ Score distributions and relative statistical uncertainties for $H \to b\bar{b}$,
    $H \to c\bar{c}$, $H\to gg$, $H\to WW^{*}$, and $H \to ZZ^{*}$ processes in the
    $\nu\bar{\nu}H$ analysis using the holistic approach trained on 100k events.
    The purple curve denotes the signal, while the colored histograms show the backgrounds.
    The red solid curve gives the statistical uncertainty versus the score cut, and
    the vertical red dashed line marks the optimal score cut. }
    \label{fig:vvH_score_and_uncertainty}
  \end{figure*}

  Using the same architecture, another model is trained on the events passing the pre-selection, with $10^5$ events to each category.
  Figure~\ref{fig:vvH_score_and_uncertainty} presents the inferred score distributions for the five Higgs decay processes. 
  Adopting the same method to the $\mu^+\mu^- H$ channel,
  we determine the optimal score threshold by maximizing the product of signal efficiency and purity.
  Applying these cuts yields projected relative statistical uncertainties of
  $0.17\%$ for $H\to b\bar{b}$, $0.99\%$ for $H\to c\bar{c}$, $0.48\%$ for $H\to gg$, $0.45\%$ for $H\to WW^{*}$ and $3.26\%$ for $H\to ZZ^{*}$.
  By implementing the categorization strategy, these uncertainties are further refined to $0.17\%$, $0.94\%$, $0.47\%$, $0.43\%$ and $2.76\%$, respectively.
  
  Our study incorporates only the Higgs-strahlung process.
  Including the $WW$-fusion process would result in an approximately 17\% increase in the cross section for the $\nu\bar{\nu}H$ final state~\cite{CEPCStudyGroup:2025kmw}.
  Taking this increase into account, and assuming for simplicity that the selection performance remains identical despite the kinematic differences introduced by the $WW$-fusion and interference effects, the relative statistical precision could be improved by roughly 9\%.
  After scaling our results, these uncertainties achieve 0.16\%, 0.86\%, 0.43\%, 0.39\%, and 2.52\%, respectively.

  \subsection{Extrapolate to Other Higgs Generation Modes}
  \label{sec:combination}
  
\begin{figure*}[htbp]
    \centering
    \includegraphics[width=\textwidth]{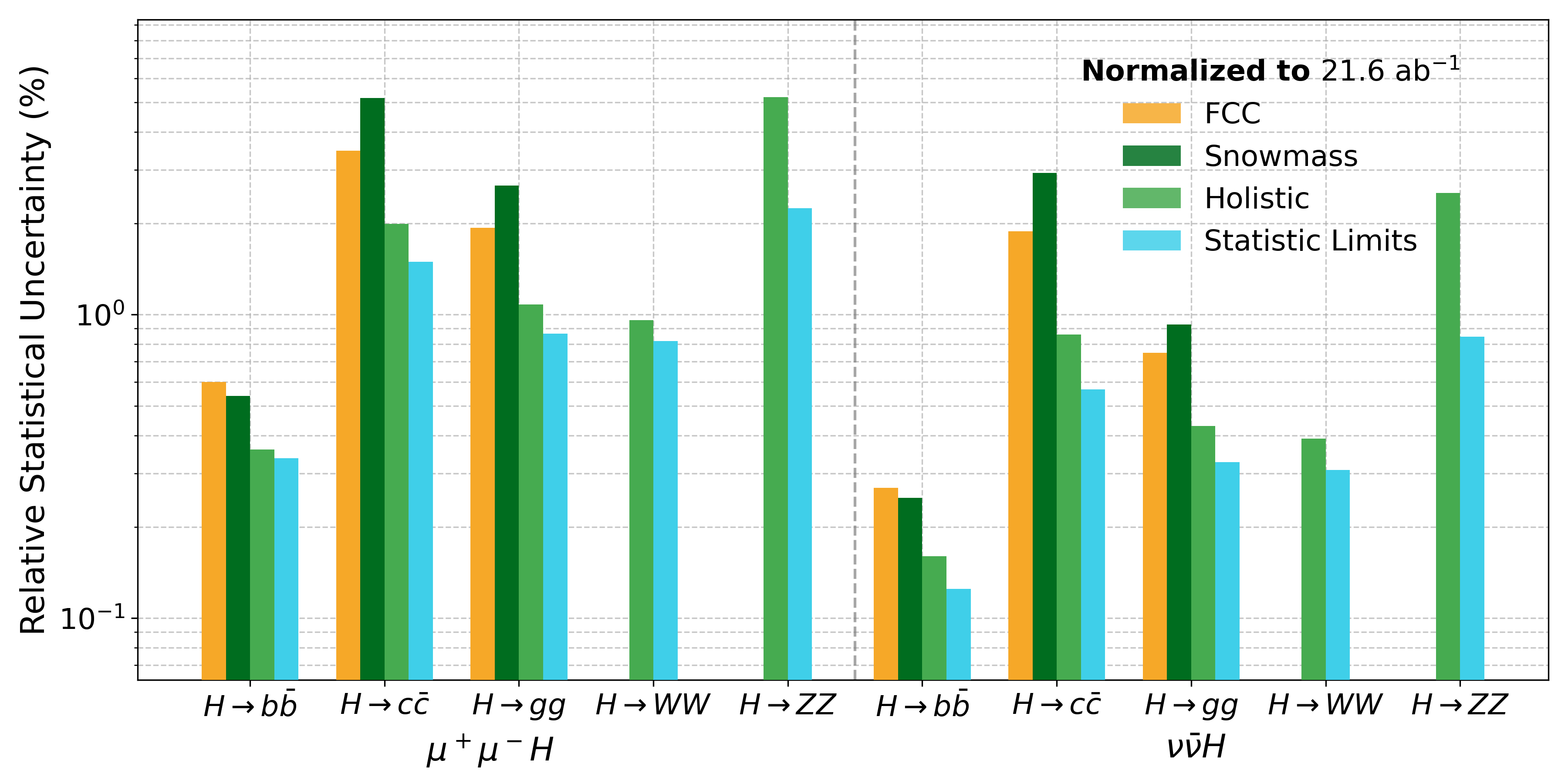}
    \caption{
    Combined precision for Higgs hadronic decay processes using the $\mu^{+}\mu^{-}H$ and $\nu\bar{\nu}H$ channels.
    The results obtained via the holistic approach are compared with the recent FCC-ee estimates~\cite{DelVecchio:2025gzw}
    and the CEPC Snowmass 2021 baseline~\cite{CEPCPhysicsStudyGroup:2022uwl},
    where the specific $H \to b\bar{b}/c\bar{c}/gg$ results are taken from Ref.~\cite{Zhu:2022lzv}.
    }
    \label{fig:combined_precision}
  \end{figure*}

  Table~\ref{tab:combined_precision} summarizes the anticipated precisions of hadronic Higgs decay modes measurements using $\mu^{+}\mu^{-}H$ and $\nu\bar{\nu}H$ channels.
  Figure~\ref{fig:combined_precision}
  compares these results with the projections from the CEPC Snowmass study~\cite{CEPCPhysicsStudyGroup:2022uwl}
  and recent FCC-ee estimates~\cite{DelVecchio:2025gzw}. For the
  $H \to b\bar{b}$, $c\bar{c}$, and $gg$ processes, the holistic approach yields significant
  improvements, reducing the relative uncertainties by a factor of approximately
  two to four compared to the Snowmass results.
  Remarkably, the precisions for these processes, together with $H \to WW^{*}$, are only 5\% to 30\% larger than the statistical limit.
  On the other hand, the $H \to ZZ^{*}$ process
  remains approximately a factor of two larger than statistical limit, as it has a relatively small branching ratio and complicated event topology (see Appendix~\ref{appendix:categorization_strategy}). 

  \begin{table}[htbp]
    \centering
    \begin{tabular}{lccccc}
      \toprule                         & $H \to b \bar{b}$     & $H \to c \bar{c}$     & $H \to g g$     & $H \to W W^*$     & $H \to Z Z^*$     \\
      \midrule $\mu^{+}\mu^{-}H$    & $0.36 \%$       & $1.99 \%$       & $1.08 \%$       & $0.96 \%$       & $5.21 \%$       \\
      $\nu \bar{\nu}H$              & $0.16 \%$       & $0.86 \%$       & $0.43 \%$       & $0.39 \%$       & $2.52 \%$       \\
      \midrule
      $e^{+}e^{-}H^{*}$             & $0.40 \%$       & $2.19 \%$       & $1.19 \%$       & $1.06 \%$       & $5.73 \%$       \\
      $\tau^{+}\tau^{-}H^{*}$       & $0.44 \%$       & $2.41 \%$       & $1.31 \%$       & $1.16 \%$       & $6.30 \%$       \\
      $q\bar{q}H^{*}$               & $0.16 \%$       & $0.86 \%$       & $0.43 \%$       & $0.39 \%$       & $2.52 \%$       \\
      \bottomrule
    \end{tabular}
    \caption{ Projected relative statistical precision for different Higgs hadronic decay processes with the integrated luminosity of $21.6~\text{ab}^{-1}$. The results for $\mu^{+}\mu^{-}H$ and $\nu \bar{\nu}H$ are derived from the holistic analysis presented in this work, with the $\nu\bar{\nu}H$ precision scaled to account for the unsimulated $WW$-fusion production. 
    The channels marked with an asterisk ($^*$) are extrapolated estimates:
    The estimation assumes $q\bar{q}H$ achieves comparable precision to
    $\nu \bar{\nu}H$, while electron and tau channels are projected from the muon
    channel assuming a 10\% cumulative performance degradation at each step.
    }
    \label{tab:combined_precision}
  \end{table}

  To provide a rough but complete prevision, Table~\ref{tab:combined_precision} also extrapolates these results to the other three Higgs production channels:
  \begin{itemize}
      \item $q\bar{q}H$ channel:
      We assume the sensitivities of those five Higgs hadronic decay modes from the $q\bar{q}H$ channel match those of the $\nu\bar{\nu}H$ channel.
      The $q\bar{q}H$ channel constitutes the majority of $ZH$ events and has a statistics that is approximately three times larger than that of the $\nu\bar{\nu}H$ events.
      Meanwhile, precisely extracting Higgs property information from $q\bar{q}H$ events requires efficient identification of the Higgs decay products, which is a challenging task in fully hadronic events.
      In classical approaches relying on jet clustering and matching, the misidentification between of Higgs and $Z$ decay products is essentially the bottleneck for these measurements. Empowered by AI, the ACSI~\cite{Zhu:2025eoe} has been developed recently, which reduces the error rate by a factor of four to five compared to the classical alternative.
      Therefore, considering the combinatorial grouping bottleneck could be solved, the hadronic measurements could match or even exceed, those of the $\nu\bar{\nu}H$ channel.
      \item $e^+e^- H$ channel: 
      We derive the projection for the $e^{+}e^{-} H$ channel from the $\mu^{+}\mu^{-}H$ baseline by assuming a 10\% degradation. On the signal side, the $e^+e^- H$ is about 7\% more abundant than that of $\mu^+\mu^-H$, as the $Z$ fusion process also leads to the $e^+e^- H$ final states; meanwhile, the background is also significantly higher than that of the $\mu^+\mu^-H$ channel, as a pattern of the electron-positron collisions. The reconstruction of $e^+e^- H$ is also less favorable compared to $\mu^+\mu^-H$ as the electrons radiate hard, while the advanced reconstruction could ameliorate these effects. Therefore, we assume a slight degrading compared to $\mu^+\mu^-H$. 
      \item $\tau^+\tau^- H$ channel: 
      We apply an additional 10\% degradation relative to the electron channel projection. The signal and background yields of $\tau^{+}\tau^{-}H^{*}$ are well comparable to $\mu^+\mu^-H$, while the identification of $\tau$ is certainly a bit more challenging than the $\mu$. A previous study shows that the $\tau$ reconstruction performance at $q\bar{q}H, H\to \tau^+\tau^-$ events (with quite similar topology to $\tau^+\tau^-H, H\to j j$) reaches an effective efficiency and purity of 80\%~\cite{Yu:2020bxh}, we estimate the performance degradation accordingly, while in fact, the $\tau$ reconstruction performance could be significantly improved by advancement in reconstruction especially those empowered by AI. 
  \end{itemize}
  
  This study ignores the interference effects for simplicity, not only in the signal channel (as the $WW$-fusion could lead to the same $\nu\bar{\nu}H$ final state interfering with $Z(\nu\bar{\nu})H$ events) but also in the background channels, where the four-fermion processes could interfere with each other. Meanwhile, as the Higgs signal could be efficiently separated from the backgrounds as demonstrated, the impact on the measurements of those leading Higgs modes  should be small.
  These results account only for the statistical uncertainties, while the control of systematic and theoretical uncertainties is critical. 
  In general, the systematic uncertainties of those Higgs measurements could, in principle, be controlled by the Z boson measurements, given the fact that CEPC could produce six orders of magnitude more Z bosons compared to the Higgs boson. The relevant systematics on detector performance (acceptance, efficiencies, differential resolution, and identification power) could, in principle, be controlled. Meanwhile, the one-one correspondence approach~\cite{Wang:2024eji} also provides a mechanism to better monitor and calibrate the detector. 
  It should be remarked that, as the holistic approach significantly improves the anticipated accuracies, the control of relevant systematic uncertainty becomes more challenging. For instance, as the $H \to b \bar{b}$ reaches per-mille level accuracy, the systematic uncertainty on luminosity measurements becomes comparable to the statistical error. Thus, the previous target of $10^{-3}$ for luminosity needs to be further improved.
  Another key component of the systematic uncertainty involves the calibration and robustness of the method itself, which is partly addressed by analyzing the scaling behavior and generator dependence in Section~\ref{sec:scaling_with_data_size}.
  While a comprehensive evaluation of systematic effects is beyond the scope of this study, we acknowledge that achieving the ultimate experimental precision will critically depend on the rigorous control of these systematics in future analyses.

  \section{Scaling behavior}
  \label{sec:scaling_with_data_size}

  We observe that better performance can be achieved by increasing the training data size, echoing the well-known scaling law at Large Language Models (LLMs)~\cite{kaplan2020scaling,hoffmann2022training}.
  We define the performance evolution as the scaling behavior, which characteristic the model’s behavior at different tasks of scientific numerical processing. 
  As characteristic pattern, the scaling behavior is then applied to optimize the analysis, to monitor the AI model, and to control the relevant systematics.

  \subsection{Signal Efficiencies and Misidentification Rates}
  \label{sec:scaling_signal_eff_misid}

\begin{figure*}[htbp]
    \centering
    \includegraphics[width=0.48\textwidth]{
      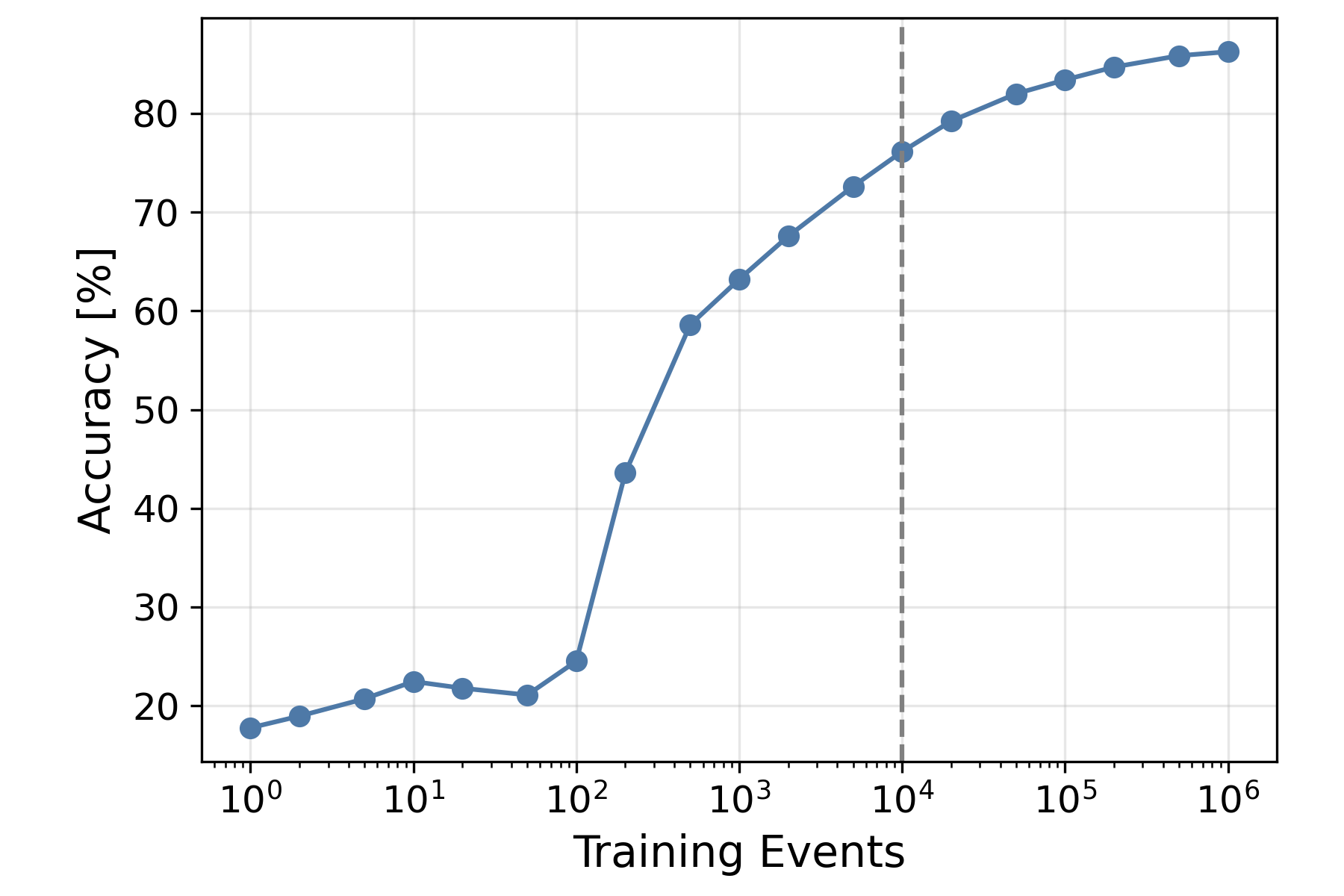
    }
    \includegraphics[width=0.48\textwidth]{
      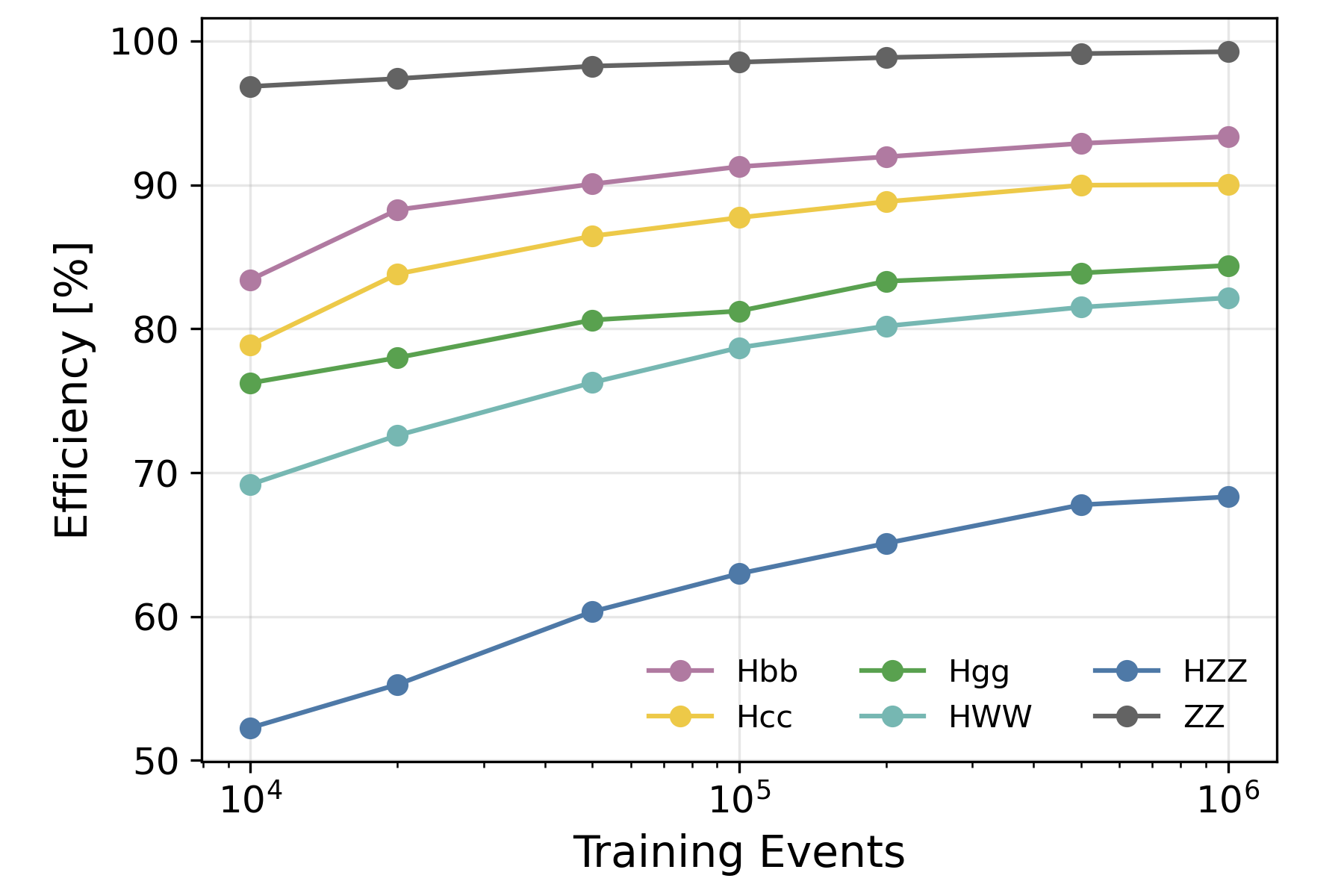
    }
    \includegraphics[width=0.48\textwidth]{
      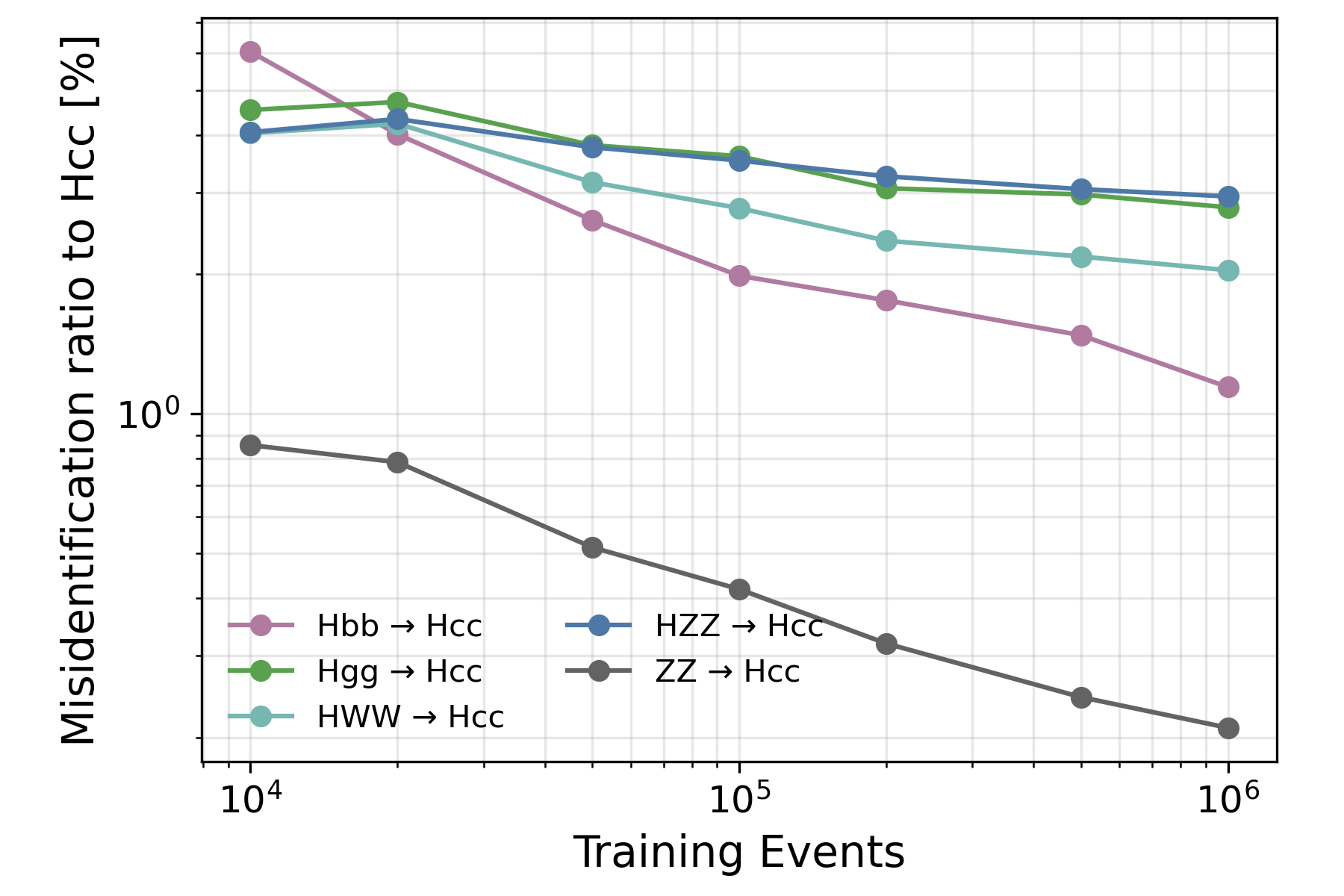
    }
    \includegraphics[width=0.48\textwidth]{
      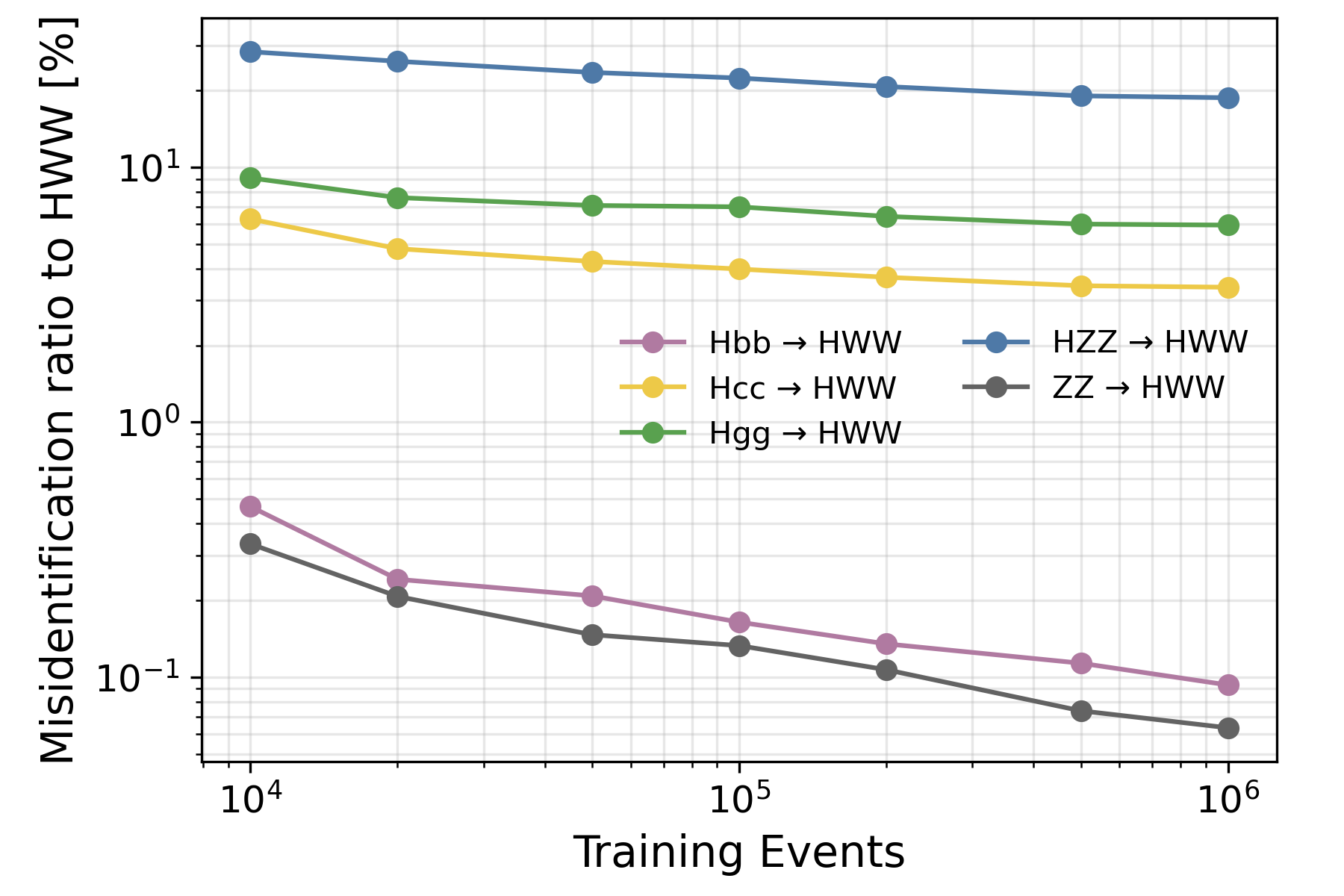
    }
    \caption{ Top left: Overall accuracy as a function of  training dataset size for the $Z(\mu^{+}\mu^{-})H$ channel. 
    Top right: Signal efficiency for each decay process versus training dataset size for the $Z(\mu^{+}\mu^{-})H$ channel. 
    Bottom left: Misidentification rate into the $H \to c \bar{c}$ channel versus training dataset size.
    Bottom right: Misidentification rate into the $H \to W W^{*}$ channel versus training dataset size. }
    \label{fig:mumuH_Overall_Accuracy_and_perclass_Acc_vs_Training_Size}
  \end{figure*}

  We evaluate the classification capability of the model as a function
  of training statistics, varying the dataset size from 1 to $10^{6}$
  events. 
  Figure~\ref{fig:mumuH_Overall_Accuracy_and_perclass_Acc_vs_Training_Size}
  illustrates the evolution of overall accuracy,
  which corresponds to the average trace of the row-normalized migration matrix,
  as well as the signal efficiencies, and specific
  misidentification rates for the $\mu^{+}\mu^{-}H$ channel.

  As shown in the top-left panel of Figure~\ref{fig:mumuH_Overall_Accuracy_and_perclass_Acc_vs_Training_Size}, the overall accuracy, defined as the average value of the diagonal elements of the migration matrix, displays a typical S-curve dependence on the training data size.
  The learning trajectory can be conceptually divided into three distinct phases:
  \begin{itemize}
      \item Trivial phase: Near-random classification until the first critical point, corresponding to roughly 100 events. 
      \item Rapid improvement phase: Performance improves rapidly with increasing data size, until roughly $10^5$ events.
      \item Slow increase phase: a slow improvement phase in which the accuracy asymptotically approaches its limit. At the maximum available statistics of $10^6$ events, the accuracy reaches approximately 86\%.
  \end{itemize}  
  
  The diagonal and off-diagonal elements of the migration matrix are extracted and presented, at training data size from $10^4$ to $10^6$.
  The signal efficiency curves (top-right panel) demonstrate distinct scaling behaviors
  across different Higgs decay processes. The vector boson decays ($H \to WW^{*}, H \to
  ZZ^{*}$) exhibit the steepest improvement slopes, indicating that larger
  datasets are particularly effective for identifying these four-jet topologies.
  In parallel, the heavy flavor modes ($H \to b\bar{b}, H \to c\bar{c}$) show
  a steady and consistent improvement, with the $H \to c\bar{c}$ accuracy increasing
  by approximately 10$\sim$15\%. Conversely, the
  $H \to gg$ channel displays a comparatively moderate improvement slope
  relative to the vector boson processes. These trends are consistent with the evolution
  of specific misidentification rates. As shown in the bottom-right panel, the misidentification
  of $H \to WW^{*}$ events as $H \to b\bar{b}$ or $H \to ZZ^{*}$ drops precipitously
  as the dataset expands. In contrast, the misidentification rates for the $H \to
  c\bar{c}$ channel (bottom-left panel) decrease more gradually, remaining the
  dominant source of background for the $c\bar{c}$ channel even at the highest
  statistics point.

  \subsection{Asymptotic Behavior of Anticipated Accuracies}
  \label{sec:asymptotic_behavior_of_anticipated_accuracies}

  Building on the classification performance, we model the dependence of the physics
  measurement uncertainty ($\delta\mu/\mu$) on the training dataset size. We employ
  a power-law ansatz with an irreducible term:
  \begin{equation}
    y = a N^{-\alpha}+ c
  \end{equation}
  \label{eq:fit}

\begin{figure*}[htbp]
    \centering
    \includegraphics[width=0.48\textwidth]{
      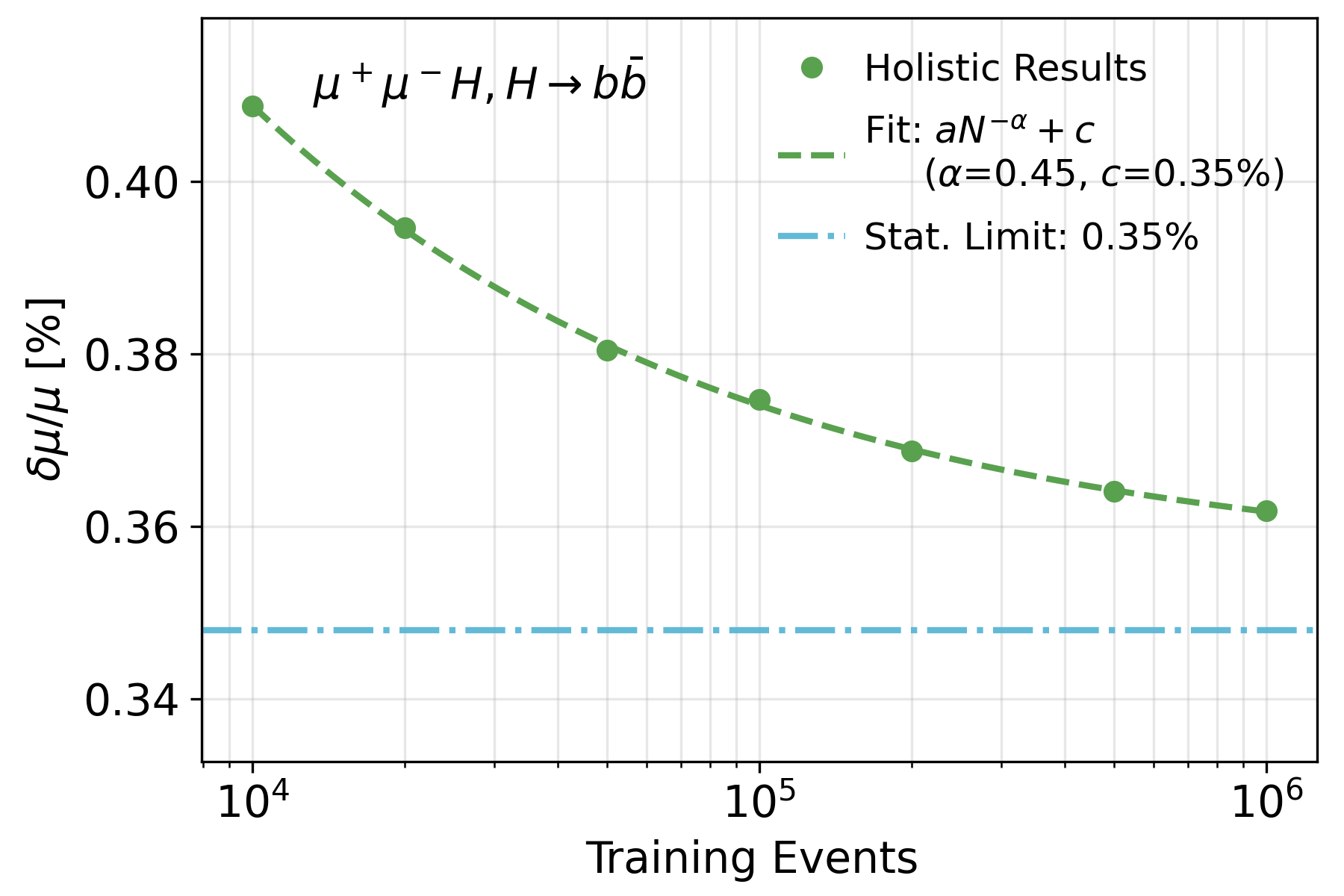
    }
    \includegraphics[width=0.48\textwidth]{
      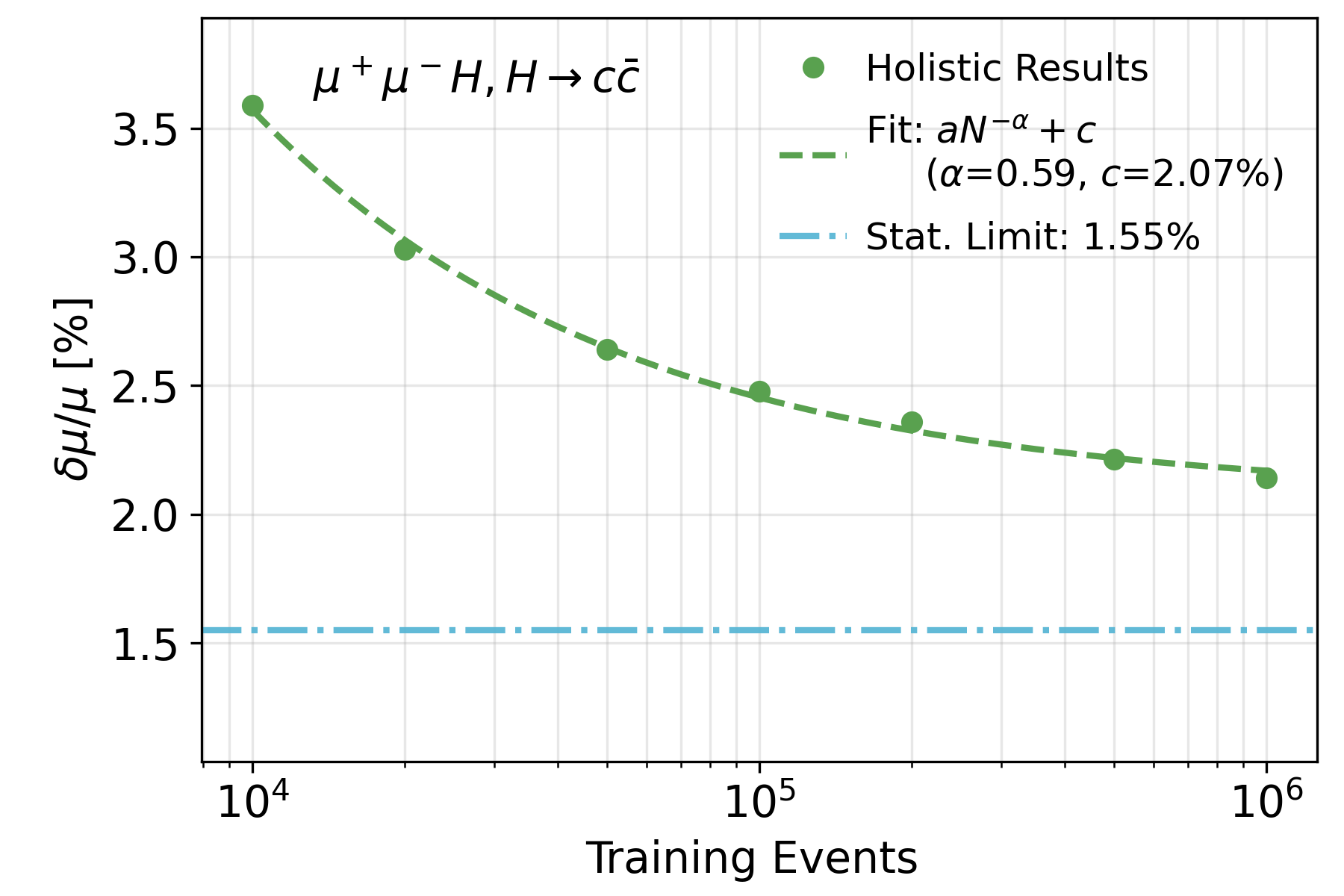
    }
    \includegraphics[width=0.48\textwidth]{
      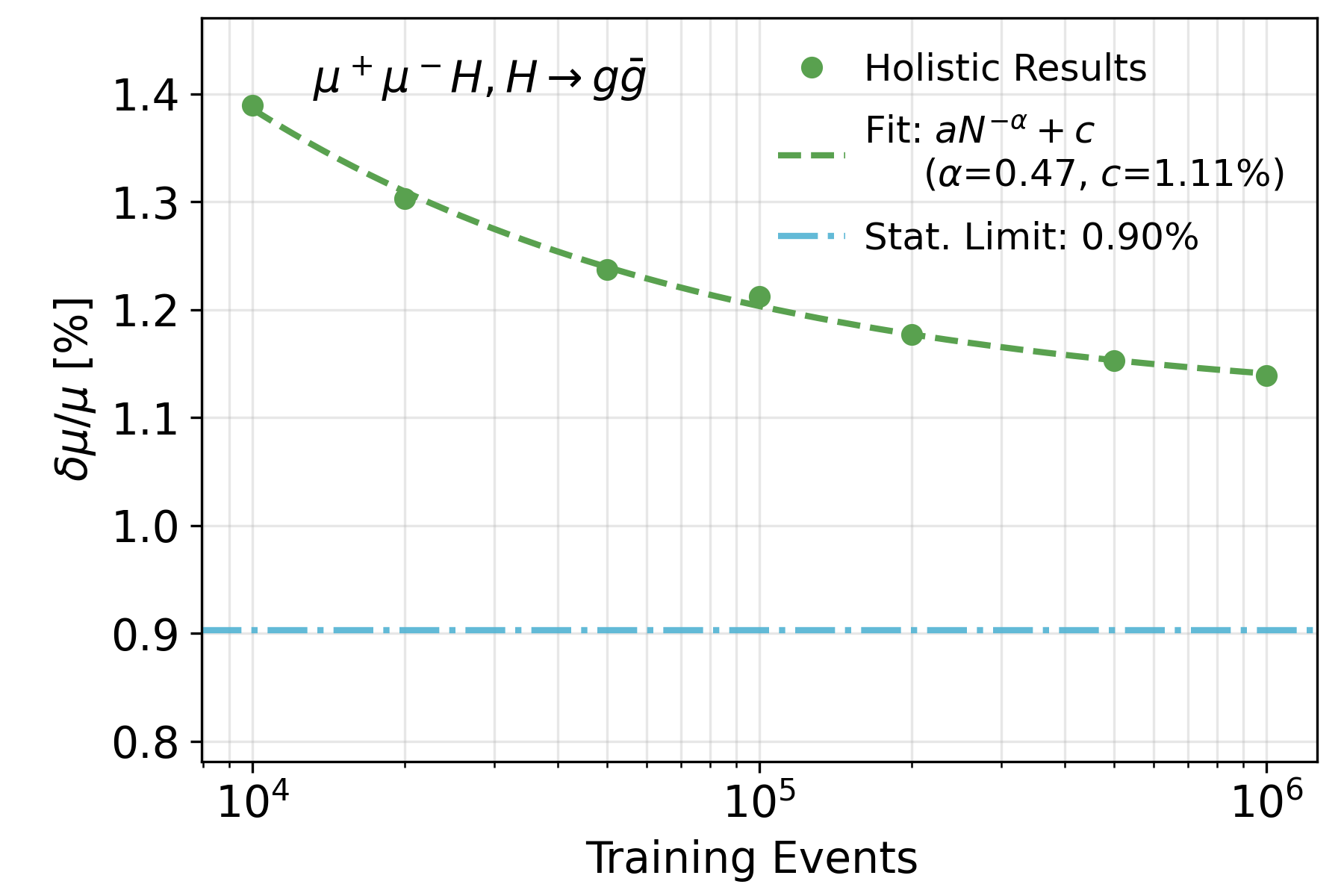
    }
    \includegraphics[width=0.48\textwidth]{
      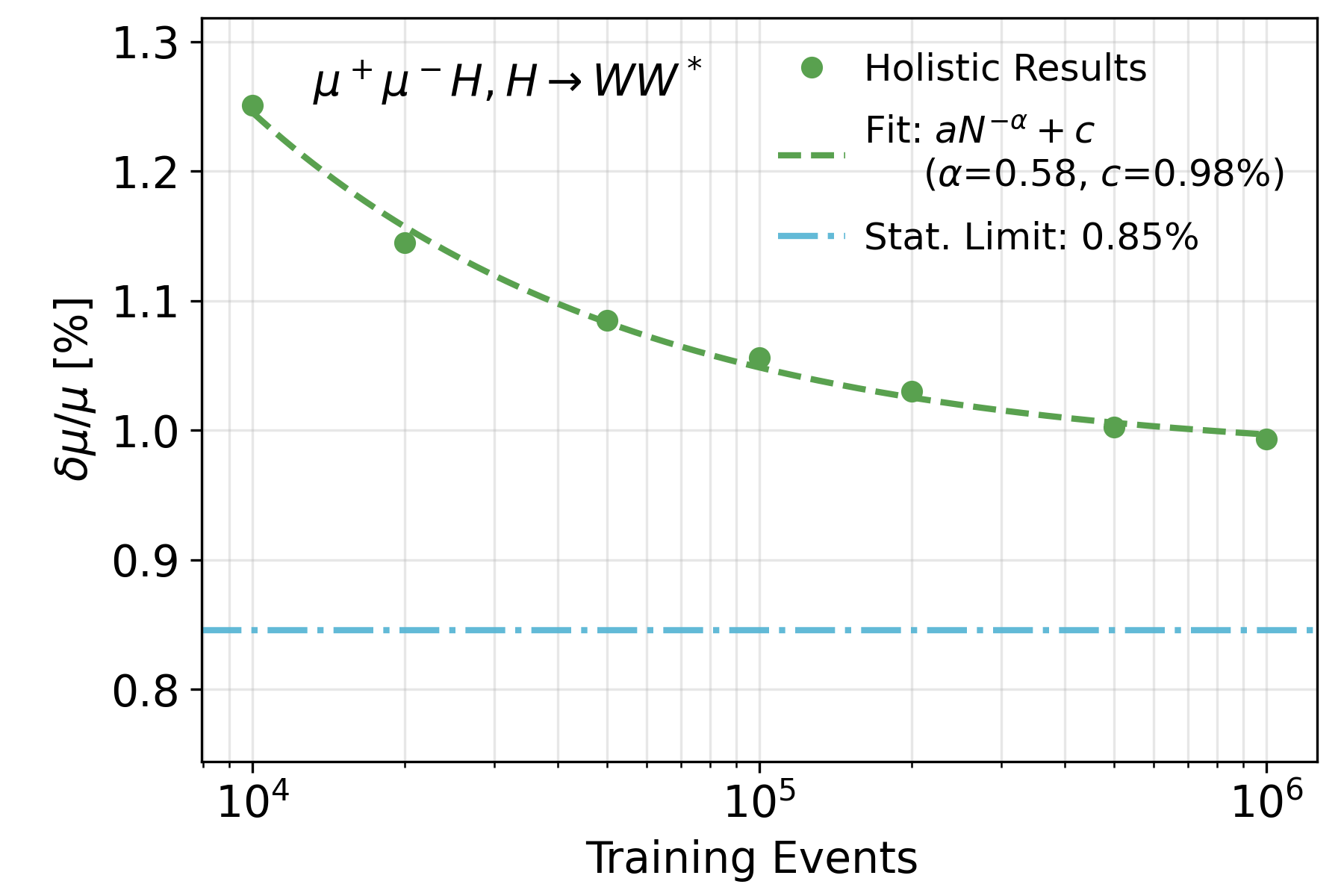
    }
    \includegraphics[width=0.48\textwidth]{
      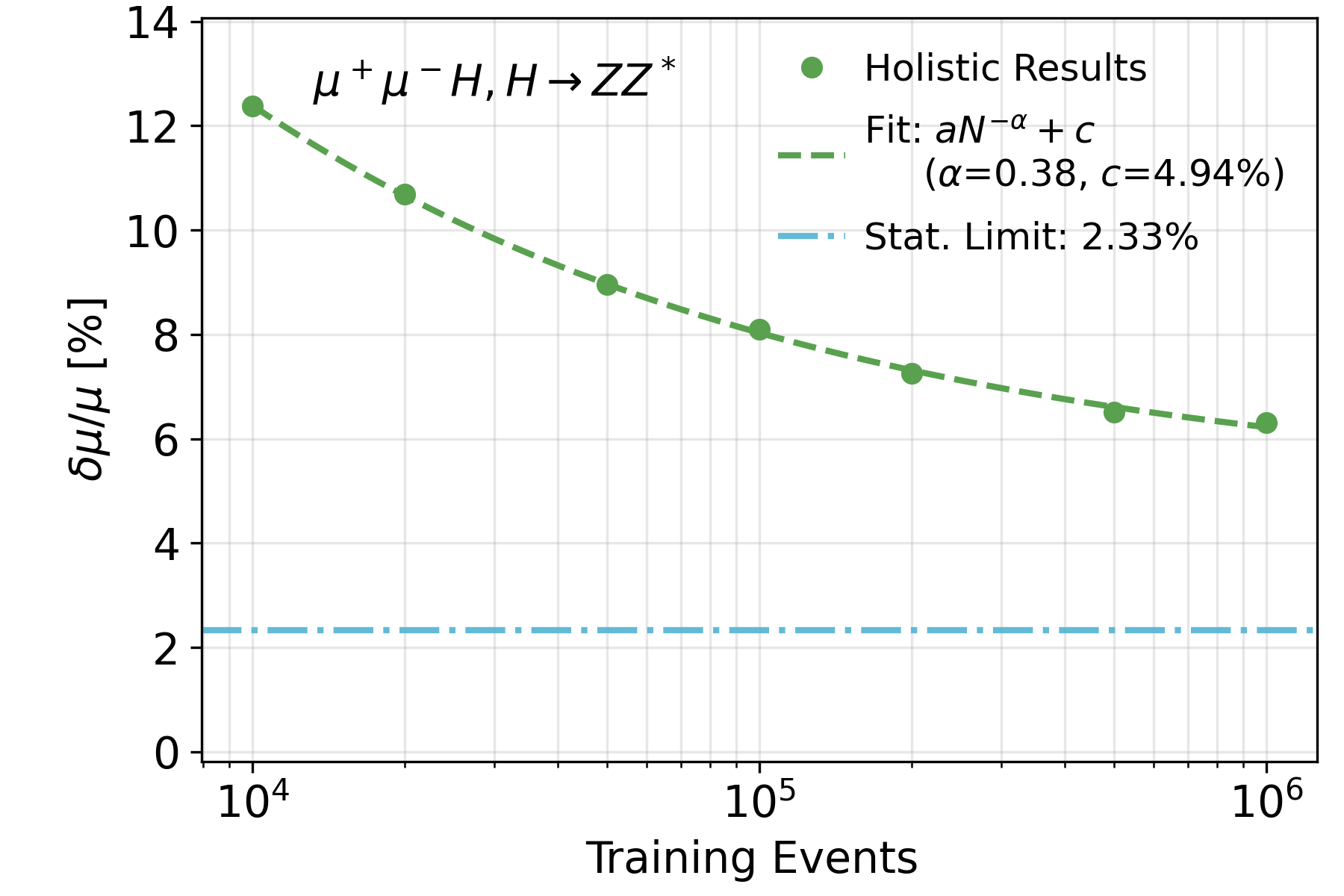
    }
    \caption{Scaling of measurement statistical uncertainty with training
    dataset size for the $Z(\mu^+\mu^-)H$ channel.}
    \label{fig:mumuH_relative_precision_scaling_laws}
  \end{figure*}

  Here, $\alpha$ denotes the scaling exponent governing the improvement rate, and $c$ represents the asymptotic uncertainty floor, which corresponds to the irreducible confusion arising from the kinematic overlap between the signal and background phase spaces.
  Applying this scaling model to the 
  $\mu^{+}\mu^{-}H$ channel, as shown in Figure~\ref{fig:mumuH_relative_precision_scaling_laws},
  the precision improves significantly with increasing statistics.

  Two different asymptotic behaviors are observed for different decay processes.
  For $H \to b\bar{b}$, the projected precision asymptotically approaches the statistical limit.
  For other processes especially $H \to ZZ^{*}$, a residual gap remains even at high statistics, suggesting the existence of a ceiling relevant to background misidentification.
  
  \subsection{Performance Dependence on Hadronization Models}
  \label{sec:generator_independence_and_robustness}

  The holistic approach utilizes reconstructed particles as input, which depend strongly on the parton fragmentation and hadronization models. 
  This dependency is also relevant to the theoretical uncertainty.
  To analyze its impact, we perform a cross-validation using
  Herwig7~\cite{Bewick:2023tfi}, which utilizes the Cluster fragmentation model~\cite{Webber:1983if}, in contrast to the Lund
  String model~\cite{Sjostrand:1982fn,Andersson:1997xwk} used by our training generator, Pythia8. We apply the models trained
  on Pythia8 datasets of varying sizes to an independent testing dataset generated
  with Herwig7.

\begin{figure*}[htbp]
    \centering
    \includegraphics[width=0.48\textwidth]{
      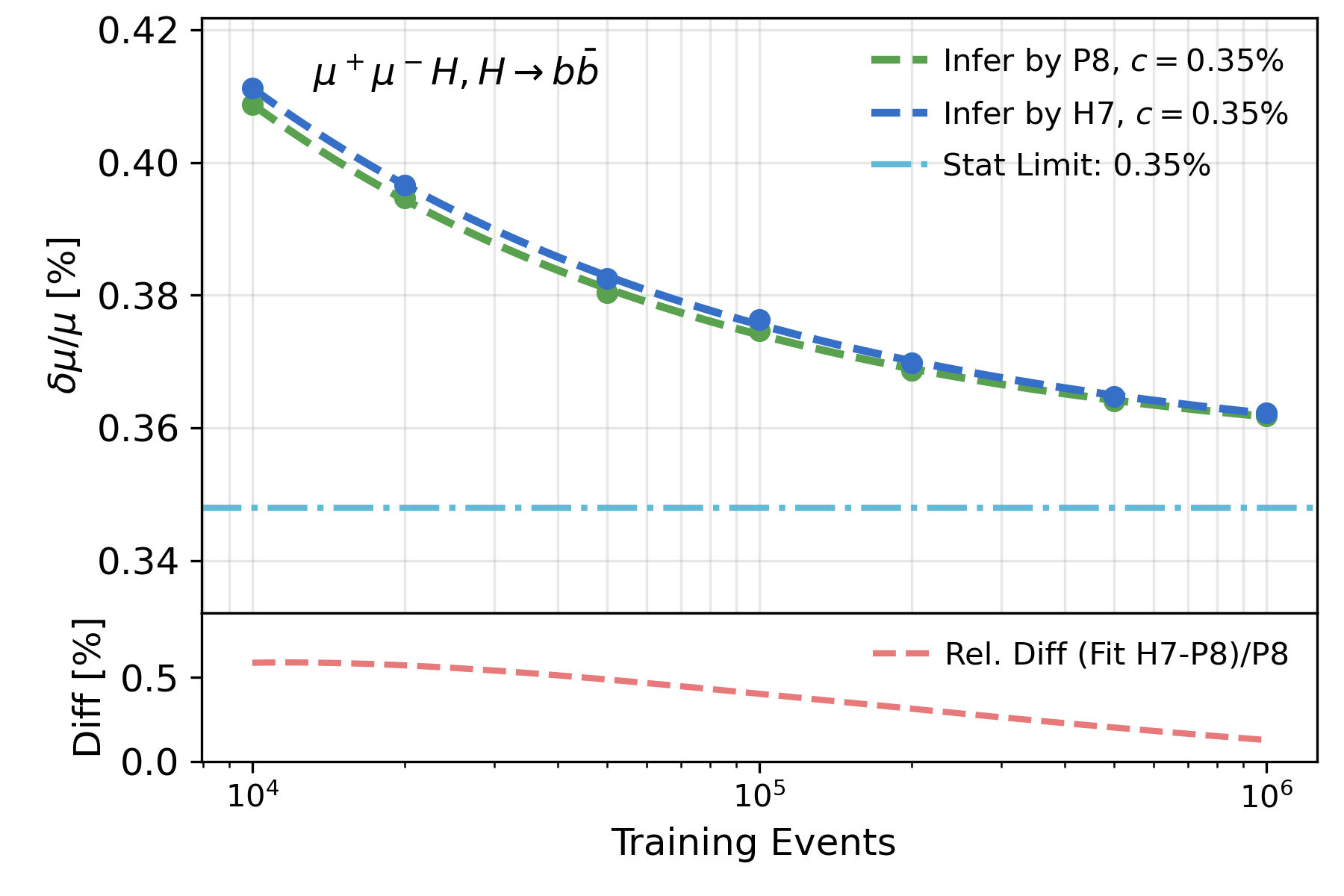
    }
    \includegraphics[width=0.48\textwidth]{
      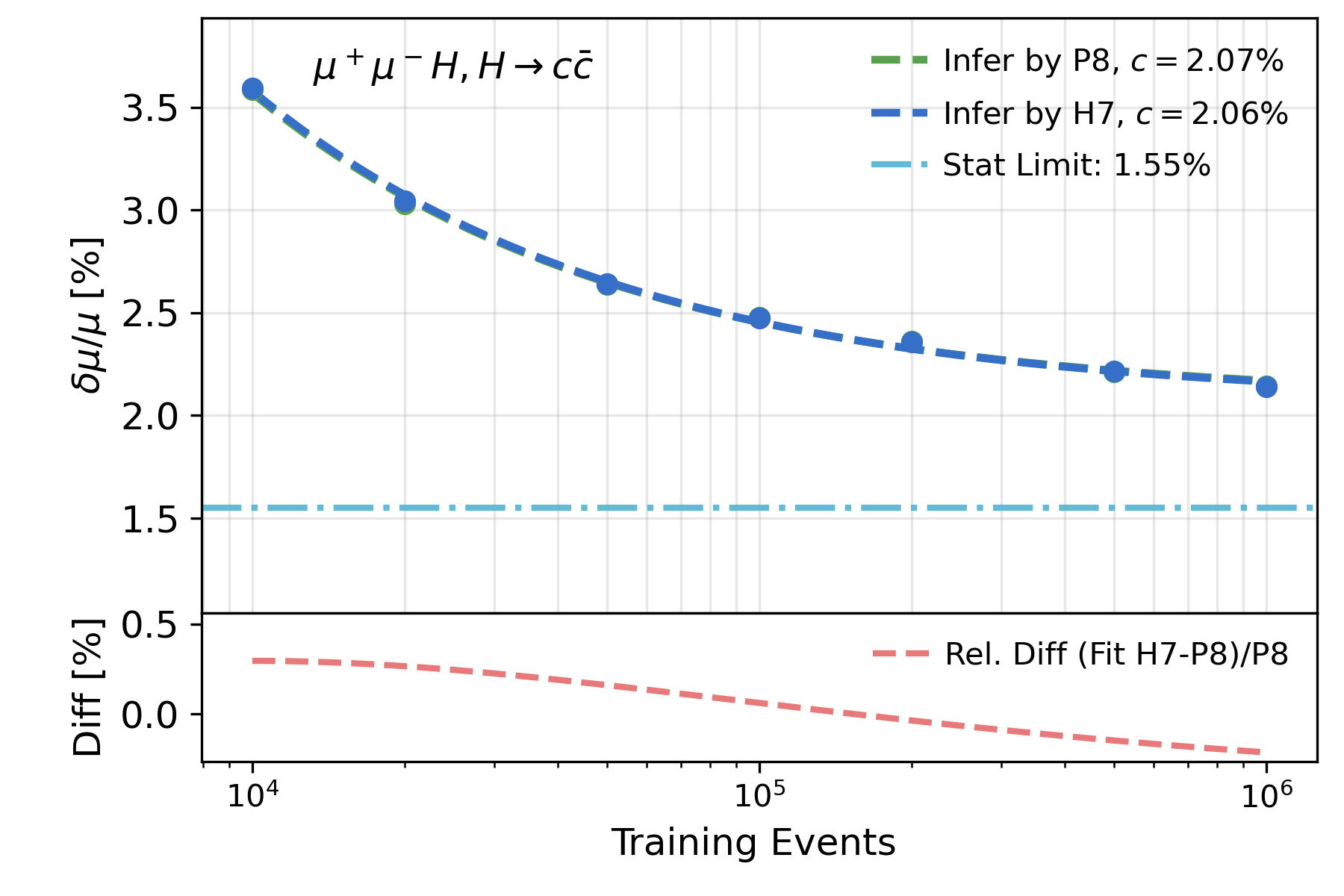
    }
    \includegraphics[width=0.48\textwidth]{
      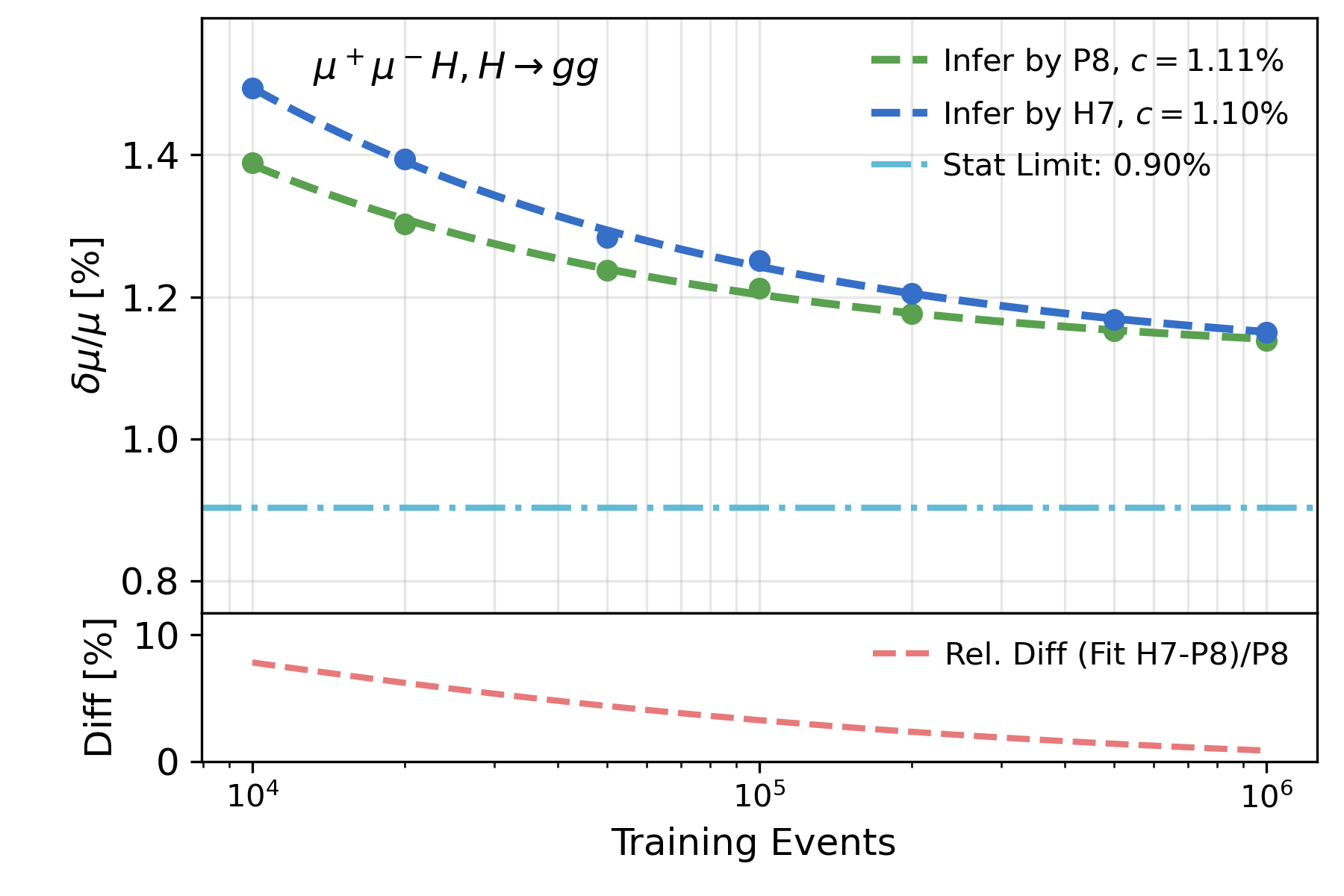
    }
    \includegraphics[width=0.48\textwidth]{
      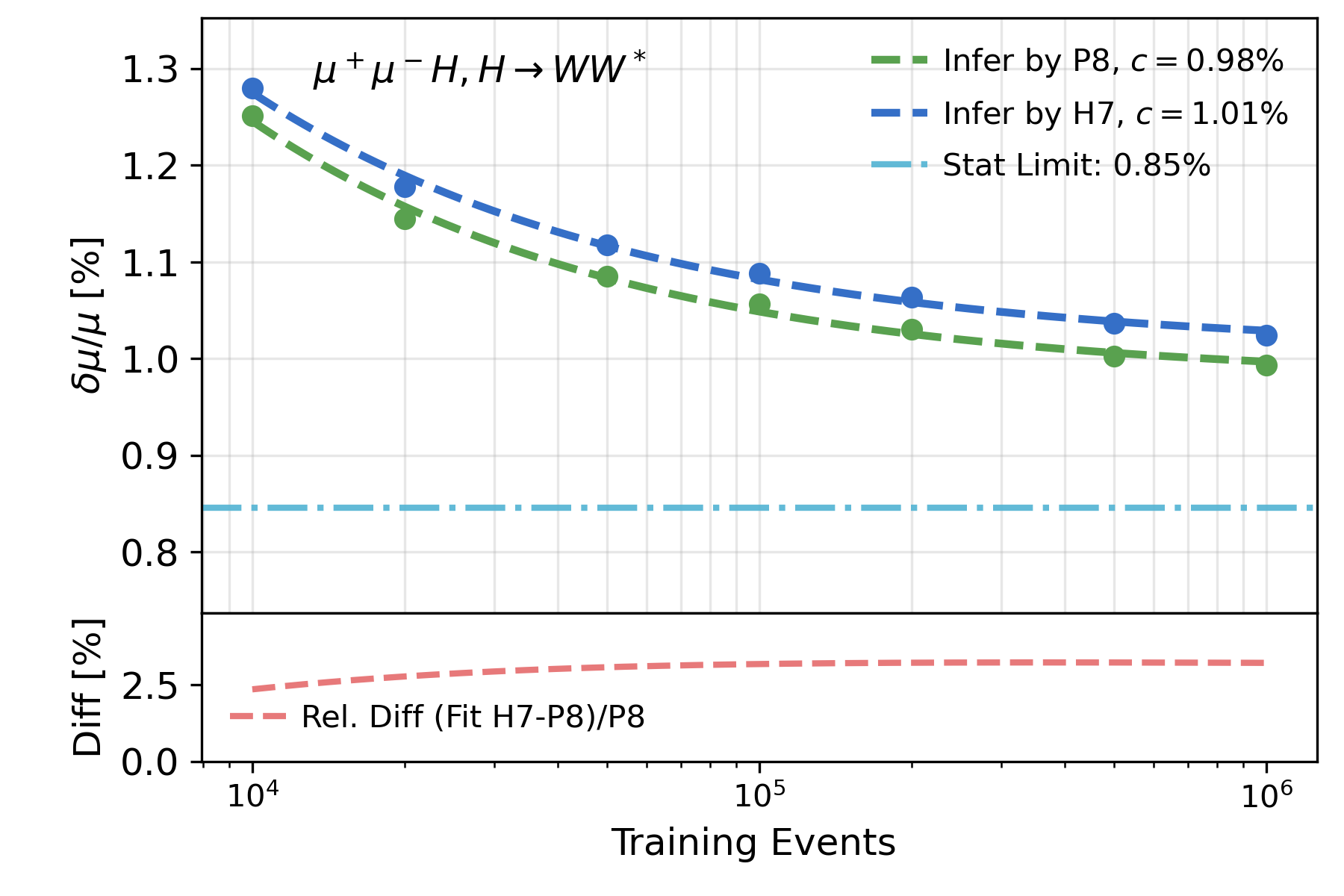
    }
    \includegraphics[width=0.48\textwidth]{
      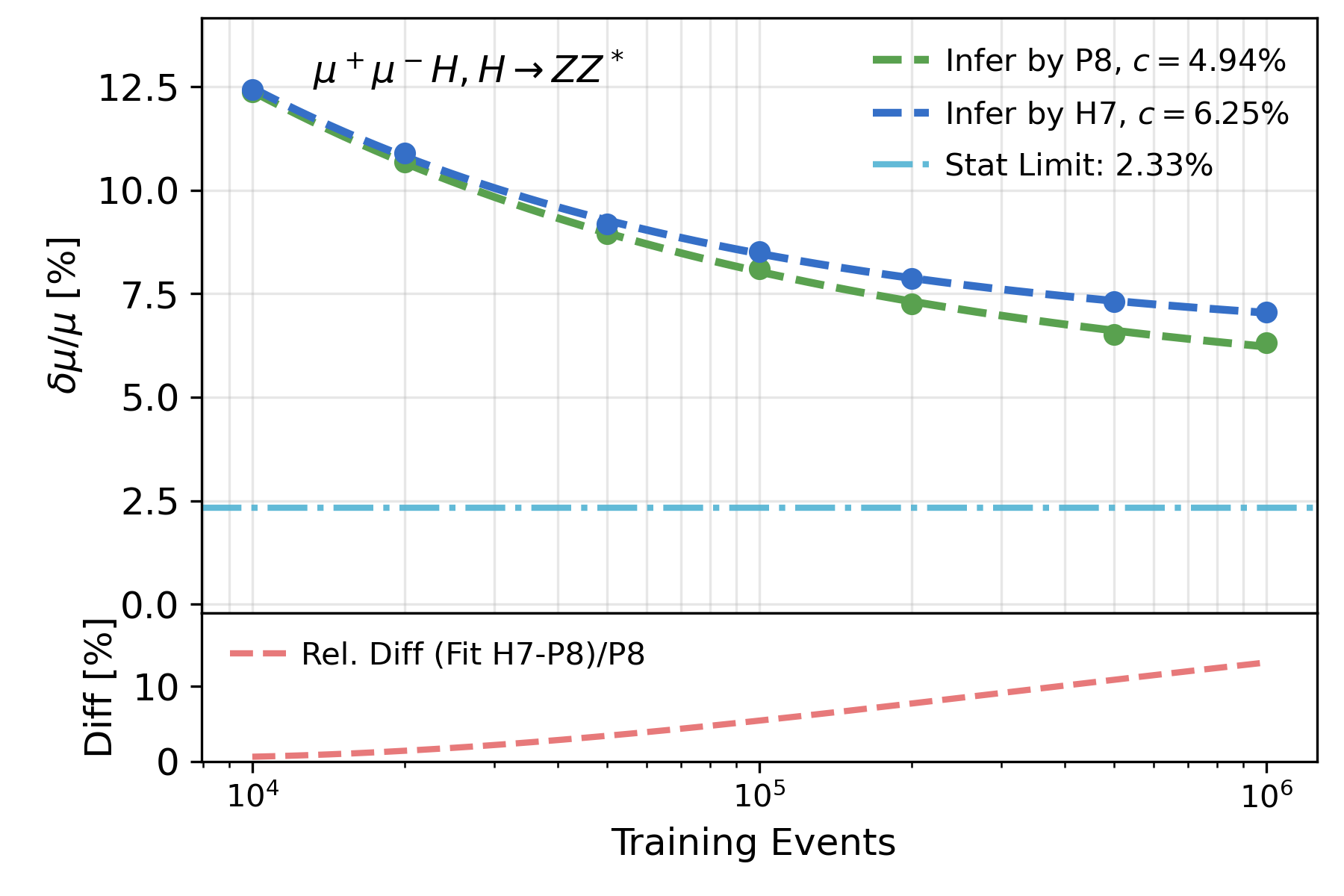
    }
    \caption{Scaling of measurement statistical uncertainty with training
    dataset size for the $Z(\mu^{+}\mu^{-})H$ channel, comparing results with Herwig7
    and Pythia8 inference. }
    \label{fig:mumuH_relative_precision_scaling_laws_compared}
  \end{figure*}

  Figure~\ref{fig:mumuH_relative_precision_scaling_laws_compared} presents the
  scaling of measurement precision for both generators. 
  The results show two
  distinct behaviors depending on the decay topology. For the two-body hadronic
  decays ($H \to b\bar{b}, H \to c\bar{c}, H \to gg$), the inference performance on Herwig7 samples
  shows a convergent trend relative to the Pythia8 baseline. As the training
  dataset expands toward $10^{6}$ events, the measurement uncertainties derived
  from Herwig7 (blue lines) asymptotically approach or maintain a stable offset
  from the Pythia8 results (green lines). This suggests that for flavor tagging
  and gluon identification, increasing the training statistics allows the network
  to focus on robust features that are consistent across the Cluster and String
  fragmentation models.
  
  In contrast, the four-body vector boson processes ($H \to WW^{*}, H \to ZZ^{*}$)
  exhibit a divergent trend. At low statistics ($10^4$ events), the performance
  gap between the two generators is minimal. However, as the statistics increase,
  the gap widens significantly: while the performance on the Pythia8 test set
  improves significantly with a steep slope, the improvement on the Herwig7 samples is
  more modest. As illustrated by the red curves representing the relative
  difference, the discrepancy increases with the training set size for these
  channels. This divergence indicates that for complex four-jet topologies, the
  network learns high-order kinematic correlations specific to the training
  generator (Pythia8) that do not fully generalize to the Herwig7 environment.
  Consequently, measurements of $H \to WW^{*}/ZZ^{*}$ will be subject to larger
  modeling systematic uncertainties compared to the two-body decay modes.

  \subsection{With/Without Pre-Selection}
  \label{sec:with_without_pre_cuts}

\begin{figure*}[htbp]
    \centering
    \includegraphics[width=0.48\textwidth]{
      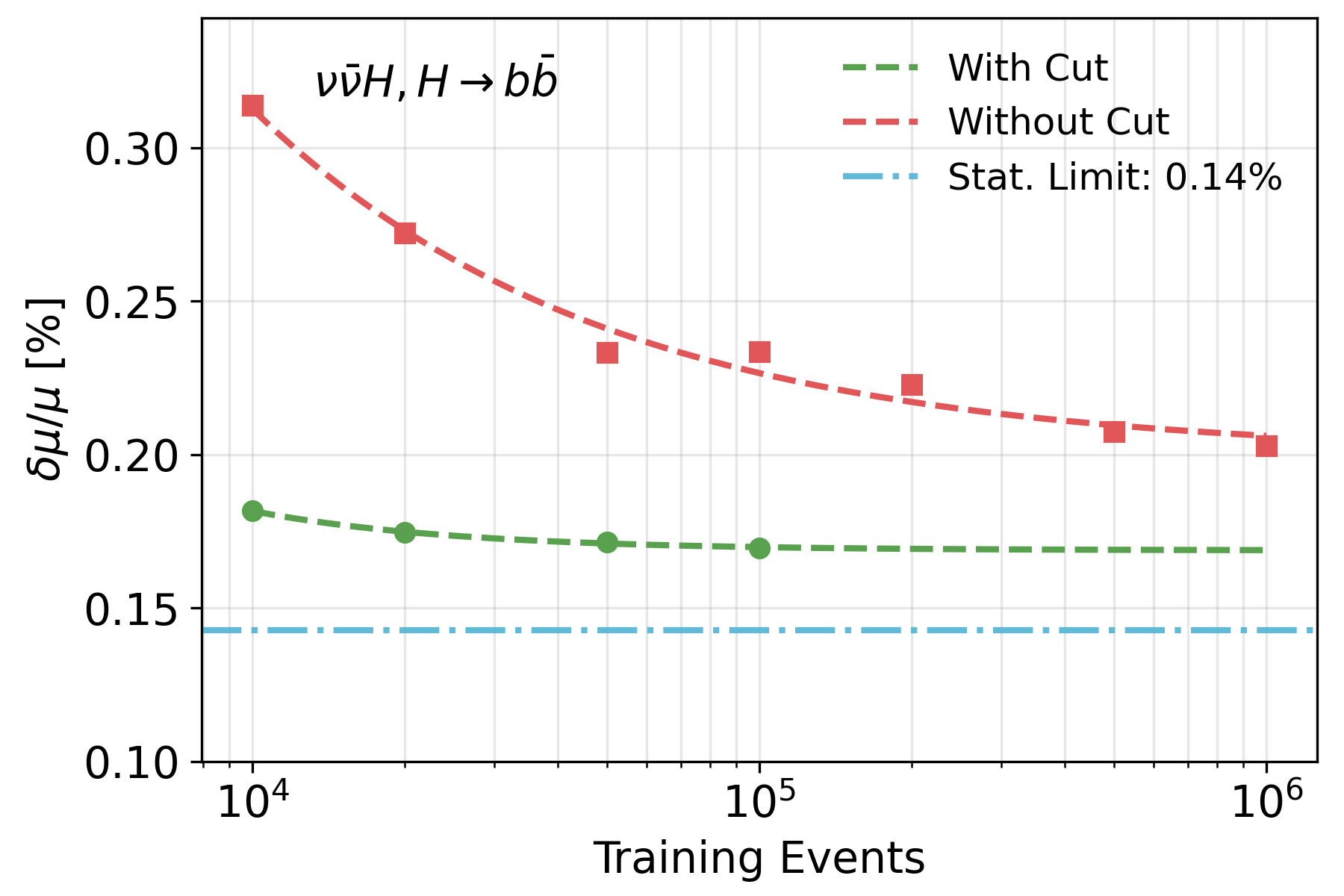
    }
    \includegraphics[width=0.48\textwidth]{
      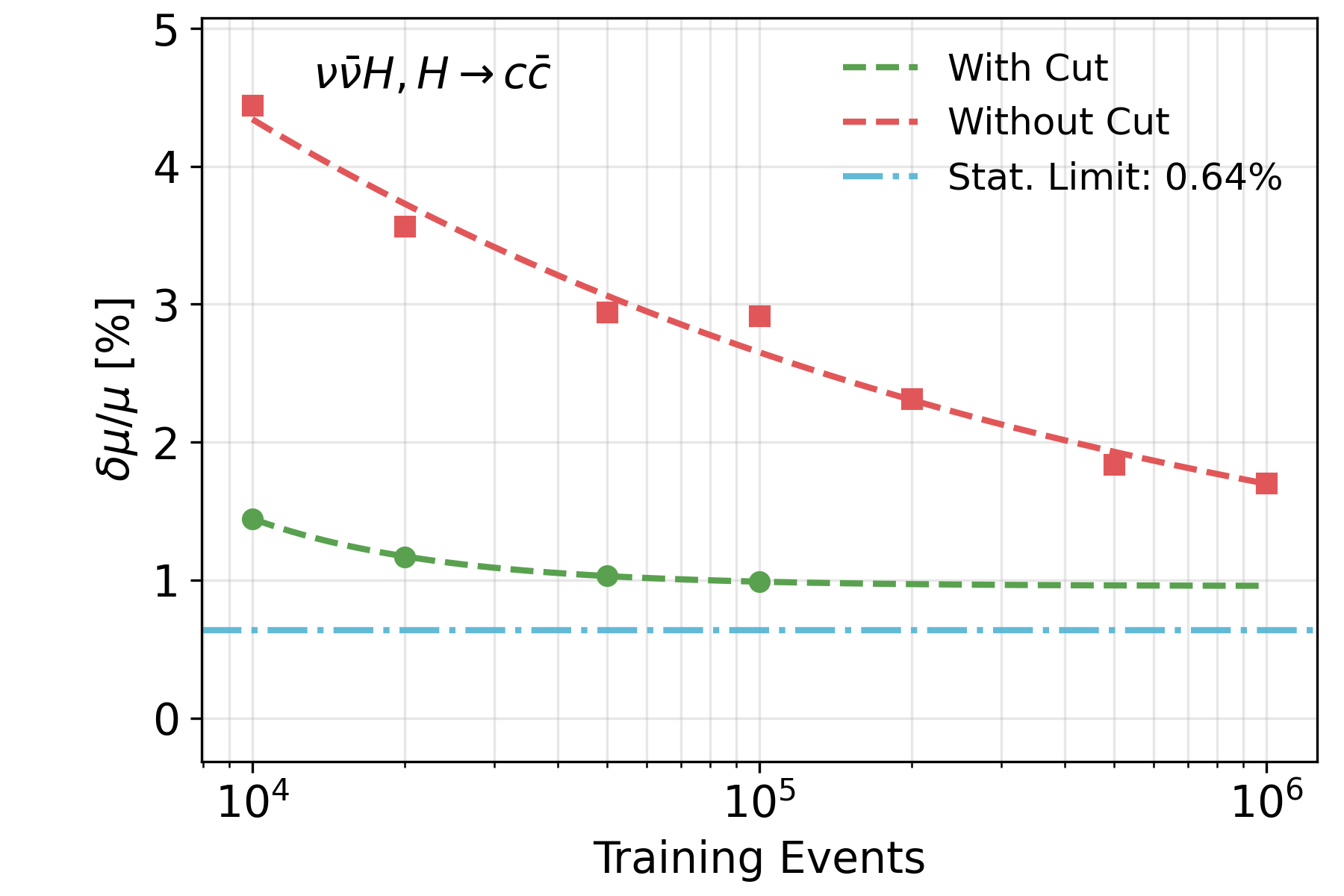
    }
    \includegraphics[width=0.48\textwidth]{
      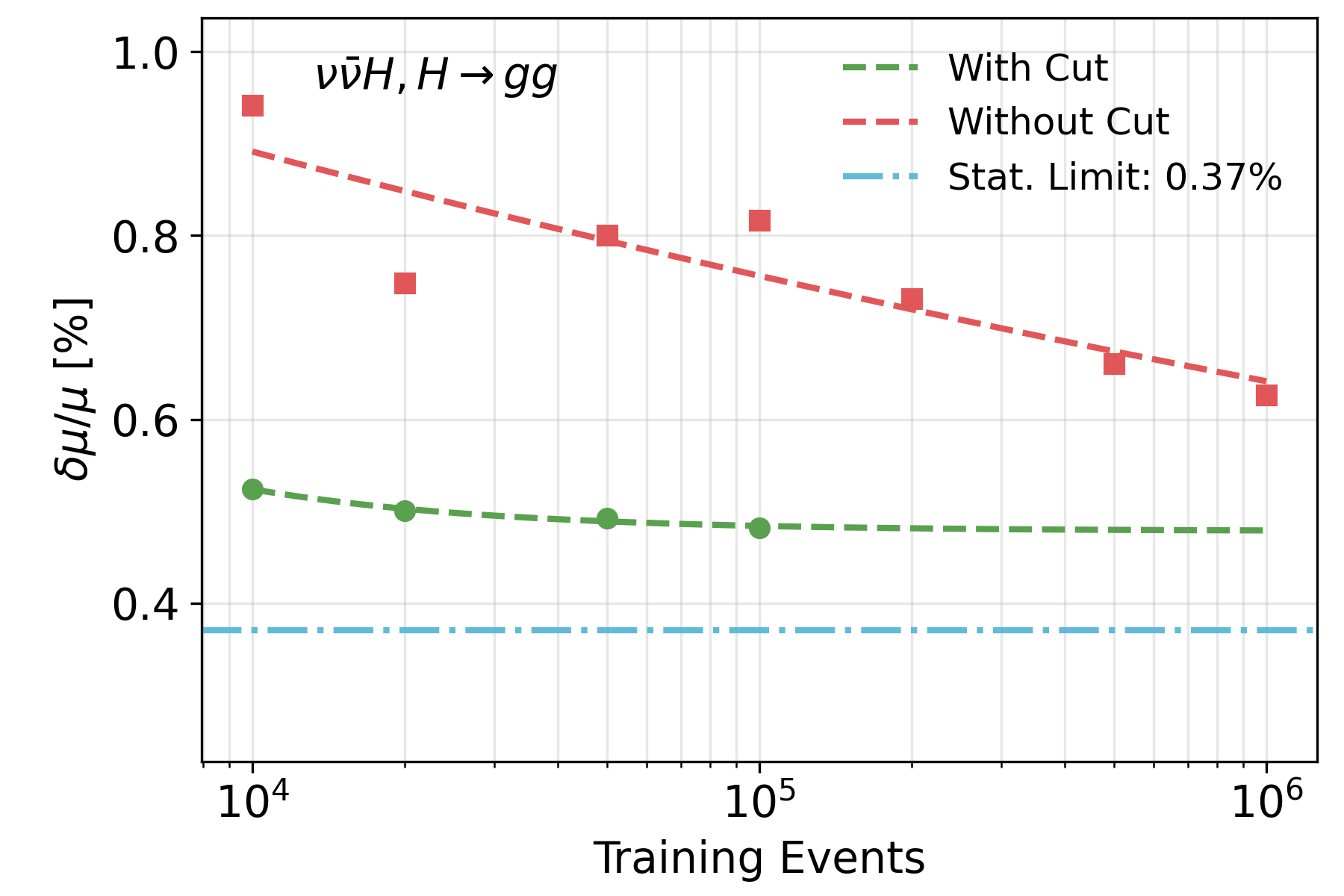
    }
    \includegraphics[width=0.48\textwidth]{
      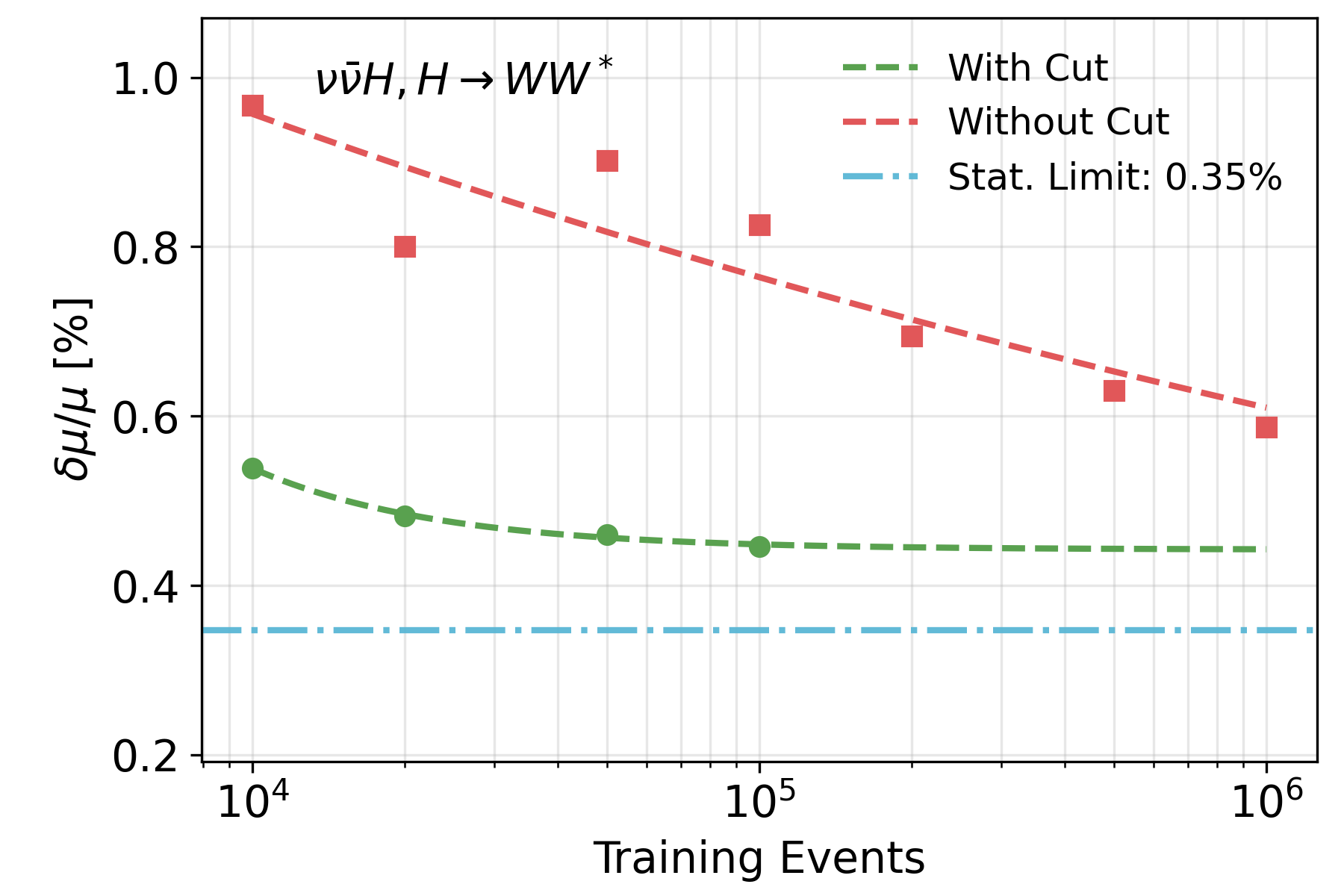
    }
    \includegraphics[width=0.48\textwidth]{
      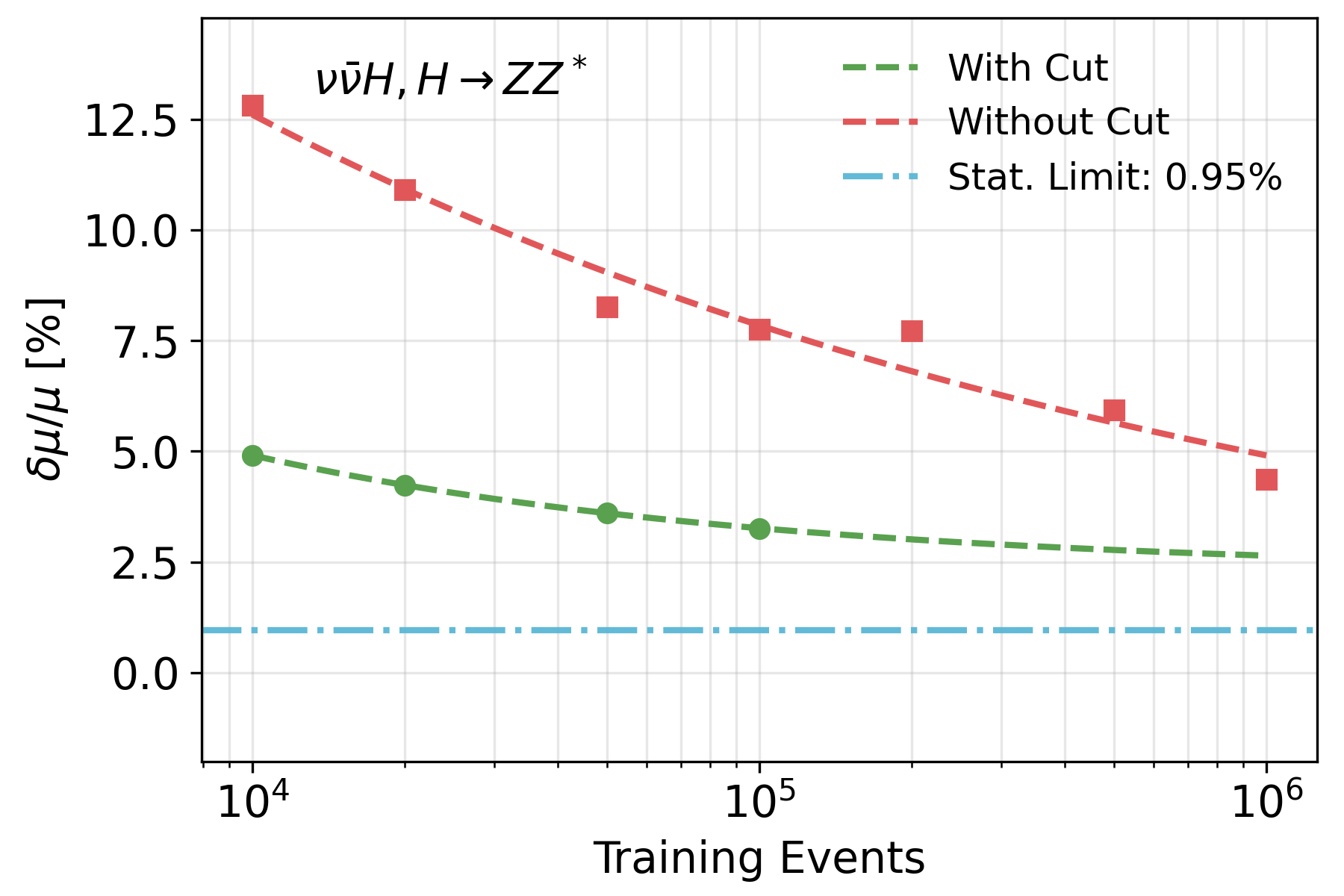
    }
    \caption{Scaling of measurement statistical uncertainty with training dataset size for the $Z(\nu\bar{\nu})H$ channel, comparing results with and without pre-selection. For the "with cut" scenario, the models are trained on datasets up to $10^5$ events, and the curves extending to $10^6$ events represent extrapolations based on the fitted scaling laws.}
    \label{fig:vvH_relative_precision_scaling_laws}
  \end{figure*}

  Figure~\ref{fig:vvH_relative_precision_scaling_laws}
  compares scaling behavior with and without pre-selection.
  It should be noted that due to the limited statistics of the surviving events after the rigorous pre-selection, the models in the "With Cut" scenario are trained on datasets of up to $10^5$ events. 
  The corresponding scaling curves are then extrapolated to $10^6$ events using the fitted scaling law (Eq.~\ref{eq:fit}) to allow for a direct asymptotic comparison with the "Without Cut" scenario, which is trained on datasets up to $10^6$ events.

  By using physics prior knowledge, pre-selection could strongly suppress the backgrounds at a certain cost to the signal efficiency, which becomes the ultimate ceiling of the final accuracies. 
  Thus, an optimal analysis tactic needs to balance the strength of pre-selection and consequent holistic approach. 
  We compare the scaling behavior of two tactics, "With Cut" and "Without Cut", corresponding to the analysis with and without pre-selection, see Figure~\ref{fig:vvH_relative_precision_scaling_laws}. 
  It shows that at small training data size, "With Cut" is significantly better than "Without Cut", while the relative difference shrinks at larger training data, while in the whole available data set (up to $10^6$), "With Cut" always has better performance than that "Without Cut". 
  In other words, in the current realization, "Without Cut" needs tremendously more data to converge, and to surpass that of "With Cut". 
  Therefore, to design better AI architecture with much faster convergence rates, and to better optimize the analysis tactic using physics prior knowledge, are critical topics to be explored in the future, especially for physics measurements with small signal and overwhelming backgrounds - those critical for discoveries.

  \section{Discussion and Summary}
  \label{sec:discussion_and_summary}

  We estimate the accuracies of Higgs hadronic decay measurements at electron positron Higgs factory using the holistic approach, which identifies the physics process of each individual event by leveraging the inclusive information of reconstructed particles.
  Realized using deep learning framework of PN, this approach yields two to four folds improvement over the traditional cut-based methods at the CEPC. 
  We analyze the scaling behavior of performance evolution with increasing size on the training data, which is then used to monitor and diagnose the deep learning performance, to optimize the analysis tactic, and to control the relevant uncertainties. 

  Applying holistic approach to the $Z(\mu^+\mu^-)H$ and $Z(\nu\bar{\nu})H$ processes with relevant Standard Model backgrounds,
  the anticipated relative statistical uncertainties for the $H\to b\bar{b}$, $c\bar{c}$, $gg$ and $WW^* \to 4q$ measurements are 5\% to 30\% larger compared to the statistical limits.
  The $H\to ZZ^* \to 4q$ measurement is more challenging; as it has relatively small branching ratio and complicated event topology,
  it is fragile to background processes, especially the irreducible ones of $H\to b\bar{b}$ and $H\to WW^*$ processes.
  By optimizing the analysis method, the anticipated accuracies of $H\to ZZ^*$ measurement can be improved by an additional 20\% beyond the holistic approach results, eventually controlling the uncertainty to within a factor of two of the statistical limit.
  In other word, while the holistic approach is powerful due to the boosted dimensionality of the input information, incorporating physical prior knowledge is still appreciated in the current realization for those challenging measurements.

  Based on the quantified accuracies of $\mu^+\mu^-H$ and $\nu\bar{\nu}H$ channels, we give a quick and rough estimation for all the other three Higgs generation channels.
  Certain simplifications were adopted to facilitate the estimation, and the corresponding limitations and relevant systematic uncertainties are discussed in Section~\ref{sec:combination}.
  It should be remarked that the presented values could still be further improved. 
  The observed scaling behavior indicate that better performance is possible with larger training datasets. While the currently anticipated accuracies, especially for the relatively more challenging channels like $H\to cc/gg/ZZ$, the improving trend is not fully saturated within current algorithm architecture. 
  Meanwhile, incorporating more information, for example using the ACSI to better grouping the reconstructed particles, shall certainly be helpful for the $H\to ZZ$ and even $H\to WW$ measurements.

  The holistic approach currently relies on the supervised learning models. 
  Its performance depends on the chosen AI architecture, the training procedure, and crucially,  the data-MC discrepancy.
  The evolution of performance with increasing training data size, known as the scaling behavior, can be used to quantify relevant impact and diagnose AI behavior.
  As detailed in Section~\ref{sec:asymptotic_behavior_of_anticipated_accuracies}, the learning trajectory follows a typical S-curve, transitioning through a trivial near-random phase, a rapid improvement phase of feature extraction, and a slow increase phase where the anticipated accuracy asymptotically approaches its statistical limit.
  This scaling behavior is also used to quantify and optimize the analysis tactic.
  Comparing the learning curves shows that while physical pre-selection accelerates convergence at low statistics, the holistic approach might achieve better accuracies when the training data is sufficiently abundant and reliable.
  Furthermore, we analyze the scaling behavior on different pairs of hadronization models. 
  As the training data statistic increases, the expected accuracies converges for different generator pairs for $H\to b\bar{b}$, $c\bar{c}$, $gg$ measurements, while still exhibit significant differences for $H\to WW^*$, $ZZ^*$ measurements.

  Leveraging the high-dimensional information contained in the input data and enabled by deep learning techniques, the holistic approach could significantly enhance the precision of key physics measurements and thus improve the discovery potential of Higgs factories.
  As an end-to-end framework, the holistic approach could in principle determine the signal and background simultaneously.
  In the current realization, however, analysis optimization guided by physics prior knowledge can still improve the performance, both in terms of output precision and computational cost.
  Another critical challenge for deep-learning-based methods is to quantify and control systematic and theoretical uncertainties, particularly those arising from data–MC discrepancies.
  In this context, the observed scaling behavior could serve as a useful diagnostic tool, providing valuable insight into the behavior of deep learning models.
   
  \section{Acknowledgements}
  \label{acknowledgements}
  We sincerely appreciate Tianji Cai for the helpful advice, and gratefully acknowledge Haijun Yang and Yanping Huang for their insightful discussions.
  This work was supported by the National Key R\&D Program of China (Grant Nos. 2024YFA1610603) and NSFC funded International Collaboration Fund for Research teams W2441004.


  \appendix
  \section{Categorization Strategy for the $H\to ZZ^*$ measurement}
  \label{appendix:categorization_strategy}
  To squeeze the measurement precision of the $H\to ZZ^*$ channel, we implement
  a categorization strategy that fully exploits the information of the holistic approach.

  In Section~\ref{sec:mmh}, event selection relies on the signal itself classification score
  to determine the optimal cut.
  But, the misidentified score distributions also carry valuable information.
  For the $H\to ZZ^*$ channel, $H\to b\bar{b}$ and $H\to WW^*$ are the primary background.
  Therefore, we investigate the distribution of the inferred $H\to b\bar{b}$ and $H\to WW^*$ scores for the $H\to ZZ^*$ events.

\begin{figure}[htbp]
      \centering
        \parbox[c]{0.59\linewidth}{
            \includegraphics[width=\linewidth]{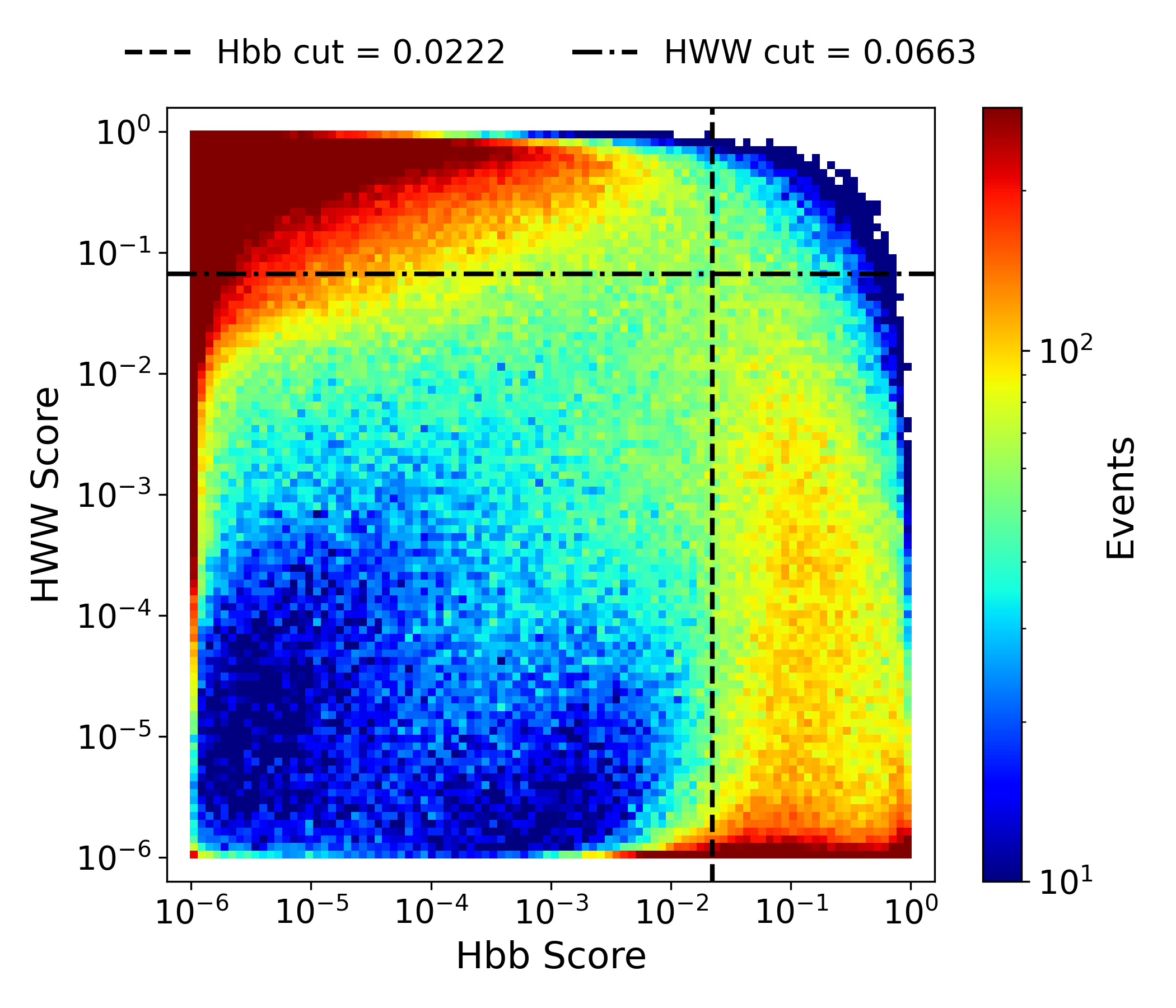}
        }
        \hfill 
        \parbox[c]{0.38\linewidth}{
        \includegraphics[width=\linewidth]{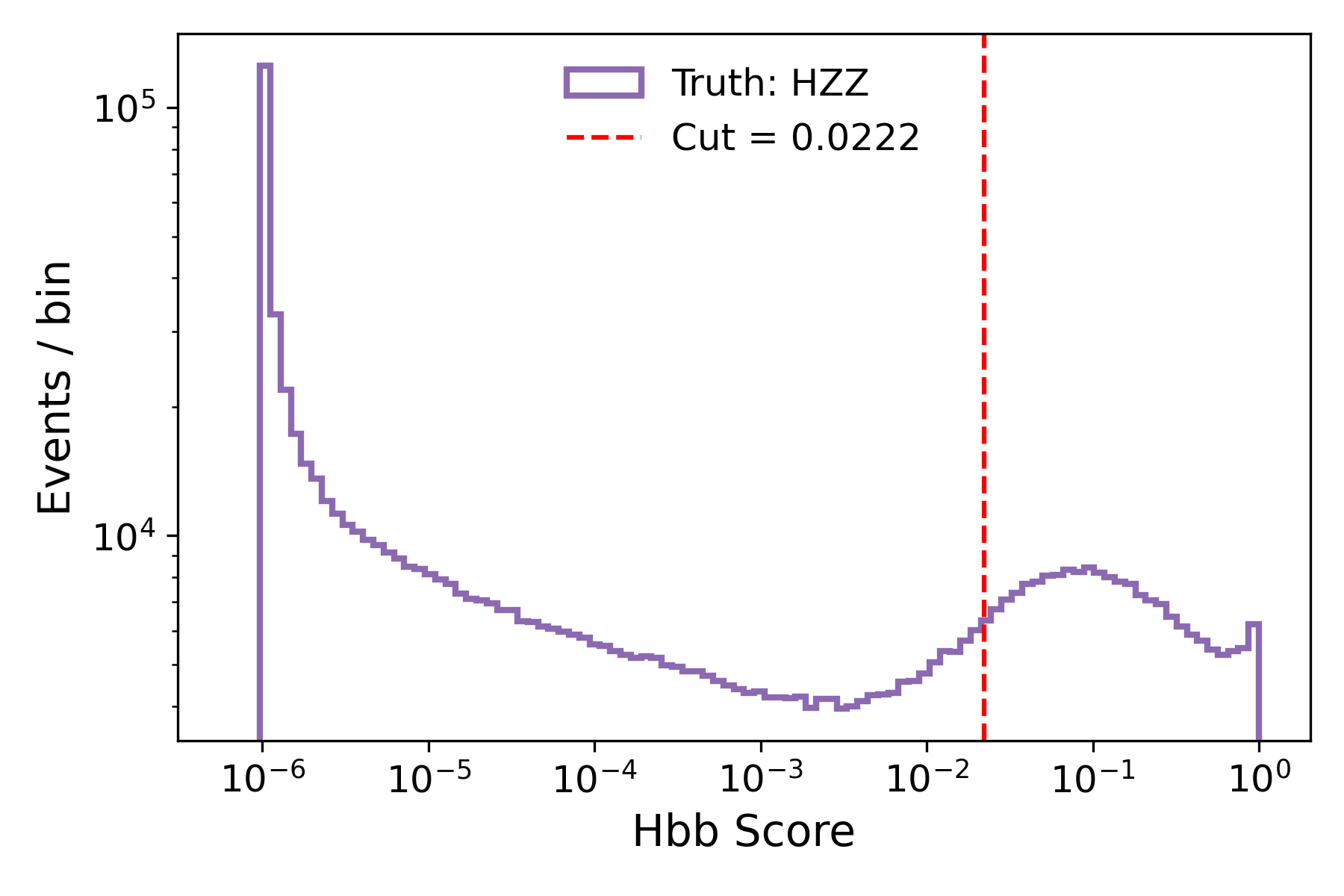}
        \\[0.05cm] 
        \includegraphics[width=\linewidth]{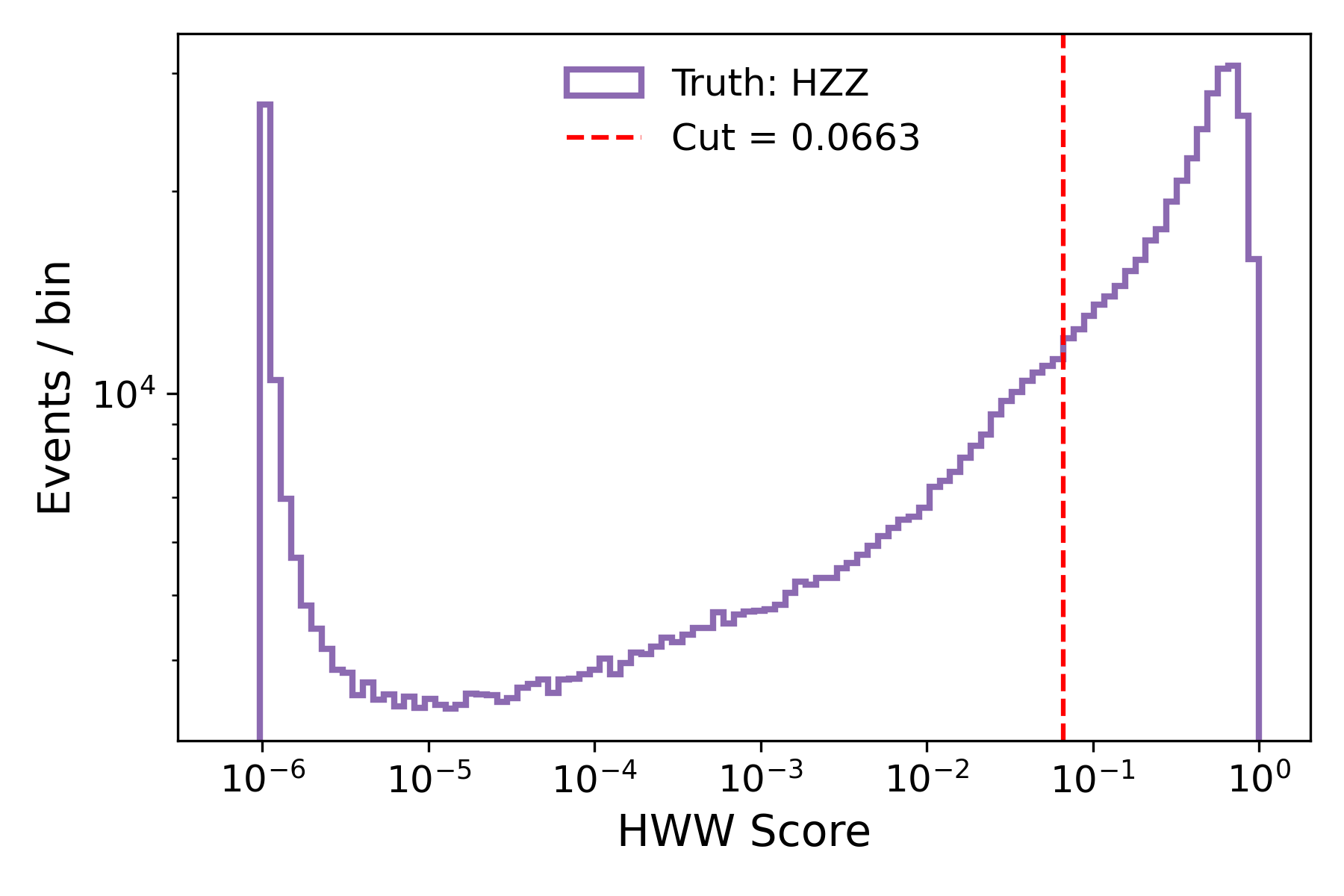}
        }
      \caption{The left side of figure shows a 2D heatmap illustrating the joint distribution of the $H\to b\bar{b}$ score and $H\to WW^*$ score for $H\to ZZ^*$ events. The dashed lines indicate the dynamically optimized splitting boundaries. The right side of figure shows the distributions of the inferred $H\to b\bar{b}$ score (top) and $H\to WW^*$ score (bottom) for the $H\to ZZ^*$ events. }
      \label{fig:HZZ_to_Hbb_HWW_2d_score_distribution}
  \end{figure}

  The distinct double-peak structures is shown in the right panel of Figure~\ref{fig:HZZ_to_Hbb_HWW_2d_score_distribution}.
  The lower score peak corresponds to events that share little topological resemblance to the respective background,
  while the higher score peak identifies events with features more akin to the background.

\begin{figure}[htbp]
      \centering
      \includegraphics[width=0.8\linewidth]{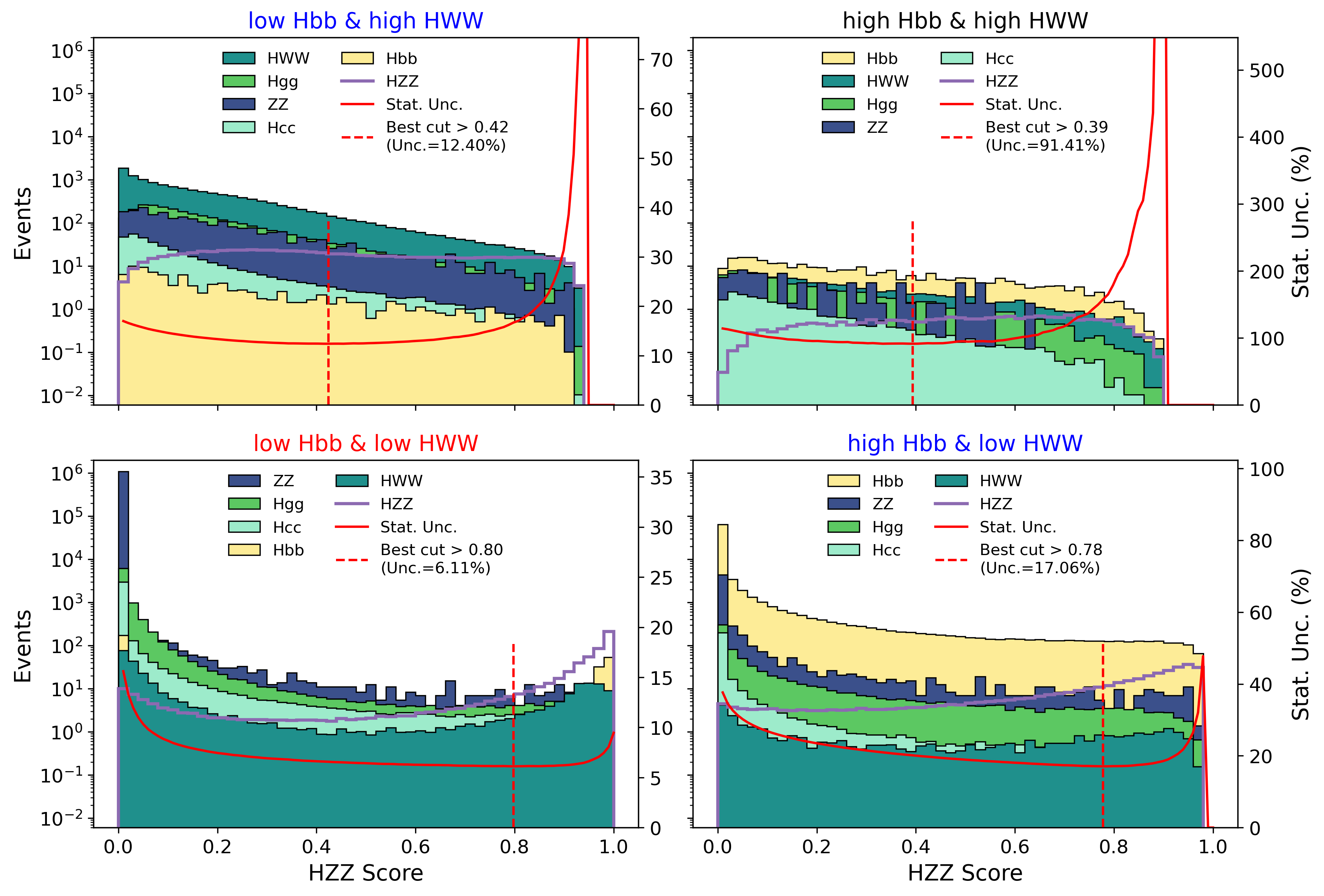}
      \caption{The $H\to ZZ^*$ score distributions and relative statistical uncertainties within the four isolated categories. The red solid curves represent the statistical uncertainty versus the score cut, and the vertical dashed lines mark the optimized thresholds for each independent region.}
      \label{fig:HZZ_to_Hbb_HWW_score_distribution_bined}
  \end{figure}

  To realize the separation, we analyze the joint 2D heatmap distribution of these two scores, shown in the left panel of Figure~\ref{fig:HZZ_to_Hbb_HWW_2d_score_distribution}.  
  We divide the phase space into four distinct categories based on a set of orthogonal cuts applied to these two score:
  \begin{enumerate}
      \item Low $H\to b\bar{b}$ and Low $H\to WW^*$
      \item Low $H\to b\bar{b}$ and High $H\to WW^*$
      \item High $H\to b\bar{b}$ and Low $H\to WW^*$
      \item High $H\to b\bar{b}$ and High $H\to WW^*$
  \end{enumerate}
  Within each of these four categories, we independently perform the same optimal cut method on the $H\to ZZ^*$ scores, as described in the Section~\ref{sec:mmh}.
  Figure~\ref{fig:HZZ_to_Hbb_HWW_score_distribution_bined} shows the score distributions and the corresponding optimal cuts for these four regions.
  The final measurement precision is obtained by statistically combining the uncertainties from these four regions.
  Because these regions are statistically independent, the combined relative statistical uncertainty ($\delta_{\text{combined}}$) is calculated using the inverse sum of variances method, yielding the formula:
  $$ \frac{1}{\delta_{\text{combined}}^2} = \sum^4_{i=1}{\frac{1}{\delta^2_i}}$$

  The exact splitting boundaries are determined dynamically by scanning the 2D score distribution to minimize the combined statistical uncertainty of the $H\to ZZ^*$ measurement.
  Applying this optimized categorization strategy yields a combined relative statistical uncertainty of 6.31\%. 
  This represents an approximate 20\% improvement over the baseline result of 5.21\% obtained from a single, global optimal cut.





  \bibliographystyle{JHEP}
  \bibliography{biblio.bib}




\end{document}